\documentclass[aps,prd,twocolumn,superscriptaddress,nofootinbib,floatfix]{revtex4-2}

\usepackage{graphicx}
\usepackage{amsmath,amssymb}
\usepackage{bm}
\usepackage{hyperref}
\usepackage{xcolor}
\usepackage{orcidlink}
\hypersetup{
    colorlinks=true,
    linkcolor=blue,
    citecolor=cyan,
}

\begin{document}

\title{Photon sphere, shadow, and polarized images of thick accretion flows around a quantum Lee-Wick black hole}

\author{Shokhzod Jumaniyozov \orcidlink{0009-0009-6254-5608}}
\email{sh.jumaniyozov@newuu.uz}
\affiliation{Kimyo International University in Tashkent, Shota Rustaveli Street 156, Tashkent 100121, Uzbekistan}

\author{Ahmad Al-Badawi\orcidlink{0000-0002-3127-3453}\vspace{0.1cm}}
\email{ahmadbadawi@ahu.edu.jo}
\affiliation{Department of Physics, Al-Hussein Bin Talal University, 71111 Ma'an, Jordan}

\author{Olimjon Toirov\orcidlink{0000-0003-0458-4878}}
\email{olimjon.toirov@tdtu.uz}
\affiliation{Tashkent State Technical University, University Street, Tashkent 100095, Uzbekistan}

\date{\today}

\begin{abstract}
We investigate the optical appearance of a static, spherically symmetric black hole (BH) arising in sixth-derivative Lee-Wick gravity, characterized by an exponential coupling parameter $S_1$ and an oscillatory coupling parameter $S_2$. Starting from the null-geodesic effective potential, we derive the photon sphere condition, the critical impact parameter, and the shadow radius of the Lee-Wick BH, and we quantify how $S_1$ and $S_2$ deform the shadow relative to the Schwarzschild case. We show that, in the weak-coupling (Block-1) regime, the shadow radius converges rapidly to the Schwarzschild value $3\sqrt{3}\,M$, while in the strong-coupling (Block-2) regime the interplay between exponential damping and oscillatory modulation produces non-monotonic, and in some regions strongly suppressed, shadow sizes. {We classify the $(S_1,S_2)$ plane according to the number of horizons and of unstable photon spheres, show that horizonless configurations still possess a critical curve but no central brightness depression, and that configurations with two unstable photon spheres produce a shadow bounded by the inner one together with an additional outer photon ring. Overlaying the 1$\sigma$/2$\sigma$ EHT shadow-size bounds of M87* and Sgr A* on the full parameter plane, we identify the excluded strong-coupling bands.} We then construct the general relativistic polarized radiative-transfer framework required to image the Lee-Wick BH when it is embedded in a geometrically \emph{thick}, optically thin accretion flow{, and implement it in a dedicated ray-tracing code that integrates the exact Lee-Wick null geodesics in their orbital planes, computes thermal-synchrotron emission at $230$~GHz from a RIAF-like flow with a ballistic velocity field, and transports the linear polarization with the Walker-Penrose constants.} We produce ray-traced shadow and polarization images across the Block-1 and Block-2 parameter sequences and as a function of observer inclination, {and compare them quantitatively with the Schwarzschild case. In the strong-coupling regime, the beam-convolved ring diameter deviates by up to $\sim14\%$, the photon-ring Lyapunov exponent grows from $\pi$ to $\simeq4$--$9$, and the amplitude of the polarimetric $\beta_2$ mode increases by up to $\sim50\%$ for a radial-toroidal field, whereas $\angle\beta_2$ is nearly insensitive to $(S_1,S_2)$ at low inclination and therefore mainly traces the field geometry.} We connect the resulting emission morphology to the shock-cone structures identified in earlier hydrodynamical studies of Bondi-Hoyle-Lyttleton accretion onto this spacetime{, showing with a parametrized shock-density model that such structures modulate the flux and brightness distribution in a strongly viewing-azimuth-dependent way while leaving the critical-curve size unchanged}, and discuss how current and future Event Horizon Telescope (EHT) observations of M87* and Sgr A* can be used to constrain $S_1$ and $S_2$. {To our knowledge, these are the first shadow, thick-disk, and polarized synthetic images of the Lee-Wick BH.}
\end{abstract}

\maketitle

\section{Introduction}
\label{sec:intro}

Higher-derivative extensions of general relativity (GR) were originally introduced to cure the ultraviolet (UV) divergences of the gravitational path integral, at the cost of introducing massive ghost-like degrees of freedom. Lee and Wick proposed that such ghosts, when arranged in complex-conjugate pairs, need not spoil unitarity of the $S$-matrix \cite{LeeWick1969,LeeWick1970,Cutkosky1969}. This mechanism, extended to gravity, yields super-renormalizable or finite quantum gravity theories \cite{Modesto2016,ModestoShapiro2018,Asorey1997,AnselmiPiva2017,LiuModestoCalcagni2023,Accioly2017,Giacchini2017}, generically referred to as Lee-Wick gravity \cite{Shapiro2023}.

An effective, spherically symmetric black-hole solution sourced by a smeared, oscillatory energy density was constructed in Ref.~\cite{BambiModestoWang2017} and has since been studied extensively: regularity and thermodynamics \cite{GiacchiniNetto2019,Burzilla2023,GiacchiniNetto2023,Netto2024}, horizon multiplicity \cite{Burzilla2023}, analogies with noncommutative-geometry-inspired black holes \cite{Nicolini2006,Nicolini2010}, gravitational lensing \cite{ZhaoXie2017,BuoninfanteGiacchini2020}, periastron precession \cite{LinDeng2022}, a rotating extension via the Newman-Janis algorithm together with its thermodynamics \cite{SinghUpadhyayAli2022,NewmanJanis1965,AzregAinou2014,DrakeSzekeres2000}, and quasinormal ringing \cite{Batic2024}. Exact solutions of the underlying sixth-derivative theory have also been explored \cite{GiacchiniKolar2024,Holdom2002}.

Most recently, the orbital dynamics of test particles and the Bondi-Hoyle-Lyttleton (BHL) accretion morphology around a non-rotating Lee-Wick BH were analyzed in detail, including the innermost stable circular orbit (ISCO), the epicyclic frequencies, and the resulting quasi-periodic oscillations (QPOs) obtained from general relativistic hydrodynamic (GRHD) simulations of the shock cone \cite{DonmezMustafaYousafJavedSaidovAtamurotov2025}. That study showed that the weak-coupling regime ($S_1 \gtrsim 2$, $S_2 \lesssim 1.5$, hereafter Block-1) reproduces Schwarzschild-like shock cones and low-frequency QPOs, whereas the strong-coupling regime ($S_1 \sim 1$, $S_2 \gtrsim 1.5$, hereafter Block-2) produces broad, asymmetric bow-shock structures and a shift toward high-frequency QPOs, with frequency ratios ($3\!:\!2$, $2\!:\!1$, $4\!:\!3$) comparable to those observed in Galactic black-hole X-ray binaries \cite{KluzniakAbramowicz2005,RemillardMcClintock2006,Pasham2018,CharlesCoe2003}.

What remains unexplored --- and what constitutes the primary aim of the present work --- is the direct \emph{horizon-scale imaging} signature of the Lee-Wick corrections $S_1,S_2$: the photon sphere, the critical curve (shadow boundary), and the polarized brightness distribution produced by a thick, optically thin accretion flow illuminating the BH. This is precisely the observable measured by the EHT for M87* \cite{EHT2019I,EHT2019V,EHT2019VI,EHT2021VII,EHT2021VIII} and Sgr A* \cite{EHT2022I,EHT2022VI}, and it has become a standard diagnostic for constraining deviations from the Kerr/Schwarzschild geometry \cite{Psaltis2020,Vagnozzi2023,PerlickTsupko2022,CunhaHerdeiro2018}. The shadow formalism traces back to the classic works of Synge \cite{Synge1966}, Bardeen \cite{Bardeen1973}, and Luminet \cite{Luminet1979}, and its use as a probe of alternative gravity and compact-object geometries is by now well established \cite{FalckeMeliaAgol2000,BambiFreese2009,HiokiMaeda2009,KumarGhosh2020,Atamurotov2015,Abdujabbarov2013,GrallaHolzWald2019,GrallaLupsasca2020,Johnson2020}.

{\emph{Relation to previous work and novelty.} The photon-sphere and shadow-radius formulas of Sec.~\ref{sec:shadow} follow the standard treatment of static, spherically symmetric metrics \cite{Synge1966,PerlickTsupko2022}, and the critical impact parameter of the Lee-Wick BH already enters, for selected couplings, the strong-deflection lensing analysis of Ref.~\cite{ZhaoXie2017}; light bending in quadratic and higher-derivative gravity was also discussed in Ref.~\cite{BuoninfanteGiacchini2020}. We do not claim novelty for these formulas. What is new in the present work is: (i) a global map of the optical structure of the Lee-Wick spacetime over the $(S_1,S_2)$ plane, including the classification of horizonless configurations and of configurations with two unstable photon spheres, and the identification of the critical curve that actually bounds the shadow; (ii) the confrontation of the full shadow-radius map with the $1\sigma$ and $2\sigma$ EHT bounds for M87* and Sgr A*; (iii) the first ray-traced intensity and polarization images of the Lee-Wick BH illuminated by a thick, optically thin 230~GHz flow, computed with a dedicated code and compared quantitatively with Schwarzschild through image-domain observables (ring diameter, brightness asymmetry, net linear polarization, and the $\beta_2$ coefficient \cite{Palumbo2020}); and (iv) an explicit, parametrized test of how shock-cone/bow-shock density structures of the type found in Ref.~\cite{DonmezMustafaYousafJavedSaidovAtamurotov2025} imprint on the images.}

In this paper we (i) derive the photon sphere and shadow-radius formulas for the Lee-Wick lapse function, present illustrative numerical results across the Block-1/Block-2 parameter space, and compare them with the Schwarzschild reference and with the angular sizes measured by the EHT for M87* and Sgr A*; and (ii) set up the general relativistic polarized radiative-transfer (GRRT) formalism for a geometrically \emph{thick}, optically thin accretion flow around the Lee-Wick BH, using both a phenomenological RIAF-like prescription \cite{NarayanYi1994,YuanNarayan2014,Broderick2011} and the analytic Hou disk model \cite{HouModel2023,ZhangThick2024}, deliberately setting aside the geometrically thin Novikov-Thorne disk. We connect the expected disk-image morphology to the shock-cone/bow-shock structures found in Ref.~\cite{DonmezMustafaYousafJavedSaidovAtamurotov2025}{, and we implement the RIAF-like framework in our own ray-tracing and polarized-transfer code (Sec.~\ref{sec:numerics}), which produces synthetic 230~GHz images and polarization maps that can be compared with EHT-like data. Public GRRT codes such as GYOTO \cite{Vincent2011}, ipole \cite{Moscibrodzka2018}, grtrans \cite{Dexter2016}, and Odyssey \cite{PuYounsi2016} were used as methodological references and for the conventions adopted, but none of them was used to generate the results presented here.}

The paper is organized as follows. Section~\ref{sec:metric} reviews the Lee-Wick BH metric. Section~\ref{sec:shadow} derives the photon sphere and shadow radius and presents numerical results{, including the classification of horizon and photon-sphere structures and the EHT constraints}. Section~\ref{sec:disk} develops the thick-disk radiative-transfer and polarization formalism {and presents the numerical implementation and the ray-traced images}. Section~\ref{sec:obs} discusses the observational strategy and EHT constraints. Section~\ref{sec:conclusions} concludes.

Throughout we use geometrized units $G=c=1$ and set the BH mass $M=1$ unless stated otherwise; $S_1,S_2$ carry dimensions of inverse length ($1/M$).

\section{Lee-Wick Black Hole Spacetime}
\label{sec:metric}

The non-rotating Lee-Wick BH is a solution of the effective field equations $G^{\mu}{}_{\nu}=8\pi G\, \widetilde{T}^{\mu}{}_{\nu}$ sourced by a smeared mass distribution with oscillatory-exponential profile \cite{BambiModestoWang2017}, arising as an approximate solution of the sixth-derivative action
\begin{equation}
S=\frac{1}{16\pi G}\int d^4x \sqrt{-g}\left[R+G^{\mu\nu}\left(S_1+S_2\Box\right)R_{\mu\nu}\right],
\end{equation}
with $S_2>0$ and $-2\sqrt{S_2}<S_1<2\sqrt{S_2}${, following the notation of Ref.~\cite{DonmezMustafaYousafJavedSaidovAtamurotov2025}}. The line element is
\begin{equation}
ds^2=-f(r)dt^2+f^{-1}(r)dr^2+r^2\left(d\theta^2+\sin^2\theta\, d\phi^2\right),
\label{eq:metric}
\end{equation}
with the lapse function
\begin{widetext}
\begin{equation}
f(r)=1-\frac{2M}{r}\left[1-\frac{e^{-rS_1}}{2S_1S_2}\Big(\big(r(S_1^2+S_2^2)S_1+S_1^2-S_2^2\big)\sin(rS_2)
+S_2\big(r(S_1^2+S_2^2)+2S_1\big)\cos(rS_2)\Big)\right] ,
\label{eq:lapse}
\end{equation}
\end{widetext}
The Schwarzschild limit is recovered for $r\gg 1/S_1$. {Near the center $f(r)\to1$ (e.g. $f(10^{-3}M)\simeq1-1.3\times10^{-6}$ for $S_1=S_2=1$), i.e. the core is regular, and $f(r)$ may even exceed unity in some intervals for large $S_2$, corresponding to a locally negative effective mass function. In what follows, and as in Ref.~\cite{DonmezMustafaYousafJavedSaidovAtamurotov2025}, the parameters $(S_1,S_2)$ appearing in Eq.~\eqref{eq:lapse} are scanned as phenomenological deformation parameters of the lapse function; part of the Block-1 range (e.g. $S_1=6$, $S_2=1$) lies outside the interval quoted above for the couplings of the action, and the corresponding results should be read as tests of the metric \eqref{eq:lapse} rather than of a specific UV completion.} Horizons are located at the zeros of $f(r)$; depending on $(S_1,S_2)$ the spacetime can have {zero, two, or four horizons over the parameter range studied here (Sec.~\ref{sec:class})} \cite{Burzilla2023,DonmezMustafaYousafJavedSaidovAtamurotov2025}. As established in the companion hydrodynamical study, two representative parameter blocks are of physical interest:
\begin{itemize}
\item[] \textbf{Block-1} (weak coupling): $S_1\gtrsim 2$, $S_2 \lesssim 1.5$ -- exponential damping dominates, $f(r)$ and the horizon structure are close to Schwarzschild;
\item[] \textbf{Block-2} (strong coupling): $S_1\sim 1$, $S_2\gtrsim 1.5$ -- the oscillatory term competes with the exponential term, producing multiple horizons or naked-singularity configurations and strongly modified curvature near $r\sim2M$.
\end{itemize}

\section{Photon Sphere and Black Hole Shadow}
\label{sec:shadow}

\begin{figure*}[tb]
\centering
\includegraphics[width=2\columnwidth]{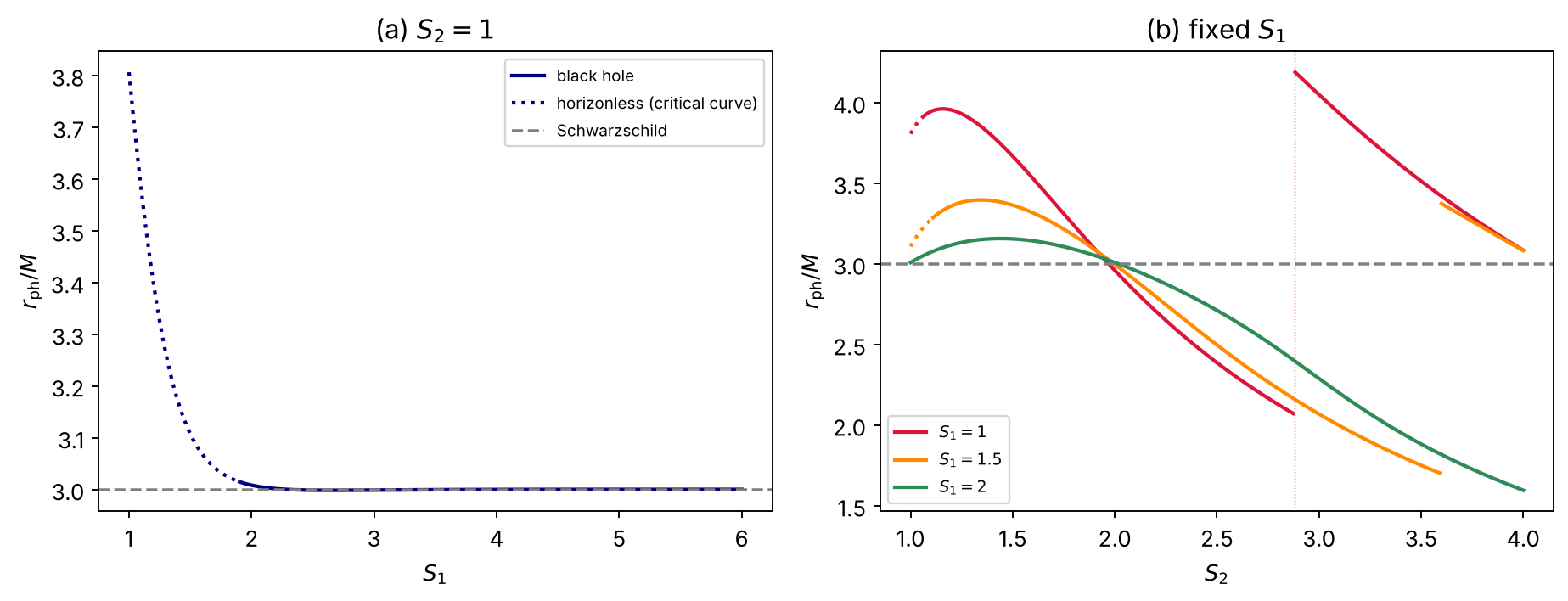}
\caption{{Radius $r_{\rm ph}$ of the shadow-bounding (innermost-relevant, see text) unstable photon sphere of the non-rotating Lee-Wick spacetime as a function of $S_1$ (left panel, $S_2=1$) and of $S_2$ (right panel, $S_1=1,1.5,2$). Dotted segments denote horizonless configurations, for which the unstable photon sphere still exists but no horizon does. Dashed thin curves in panel (b) give the radius of a second, outer unstable photon sphere whenever it exists. The dotted vertical line at $S_2\simeq2.879$ marks the horizon bifurcation of the $S_1=1$ sequence discussed in the text.}}
\label{fig:rph}
\end{figure*}

\subsection{Null geodesics and the photon sphere}

For a photon moving in the equatorial plane ($\theta=\pi/2$) of the metric \eqref{eq:metric}, the conserved energy $E=f(r)\dot t$ and angular momentum $L=r^2\dot\phi$ combine with the null condition $g_{\mu\nu}\dot x^\mu \dot x^\nu=0$ to give the radial equation
\begin{equation}
\dot r^2 = E^2 - \frac{L^2}{r^2}f(r) \equiv E^2 - V_{\rm ph}(r),
\qquad
V_{\rm ph}(r)=\frac{L^2}{r^2}f(r).
\end{equation}
The circular photon orbit (photon sphere) at $r=r_{\rm ph}$ satisfies $V_{\rm ph}'(r_{\rm ph})=0$, i.e.,
\begin{equation}
2f(r_{\rm ph})-r_{\rm ph}f'(r_{\rm ph})=0 .
\label{eq:photonsphere}
\end{equation}
For Schwarzschild, Eq.~\eqref{eq:photonsphere} gives the familiar $r_{\rm ph}=3M$. Because $f(r)$ in Eq.~\eqref{eq:lapse} is transcendental in $r$, Eq.~\eqref{eq:photonsphere} must be solved numerically for general $(S_1,S_2)$.

{Each root of Eq.~\eqref{eq:photonsphere} outside the outer horizon (or anywhere, for horizonless configurations) is either an unstable photon sphere ($V_{\rm ph}''<0$) or a stable one ($V_{\rm ph}''>0$); stable photon spheres always come paired with an additional unstable one \cite{Cunha2017}.} Each unstable photon sphere defines a critical impact parameter
\begin{equation}
b_{\rm ph}\equiv \frac{L}{E}\bigg|_{\rm crit} = \frac{r_{\rm ph}}{\sqrt{f(r_{\rm ph})}}.
\label{eq:bph}
\end{equation}
{and a Lyapunov exponent, which for metrics with $g_{tt}g_{rr}=-1$ takes the form \cite{Cardoso2009}
\begin{equation}
\gamma=\pi\sqrt{\frac{2f(r_{\rm ph})-r_{\rm ph}^2f''(r_{\rm ph})}{2}},
\label{eq:lyap}
\end{equation}
with $\gamma=\pi$ for Schwarzschild. The quantity $e^{-\gamma}$ sets the flux ratio of successive higher-order (``$n=1,2,\dots$'') photon-ring sub-images \cite{Johnson2020}.}

\subsection{Shadow radius and celestial coordinates}

\begin{figure*}[tb]
\centering
\includegraphics[width=2\columnwidth]{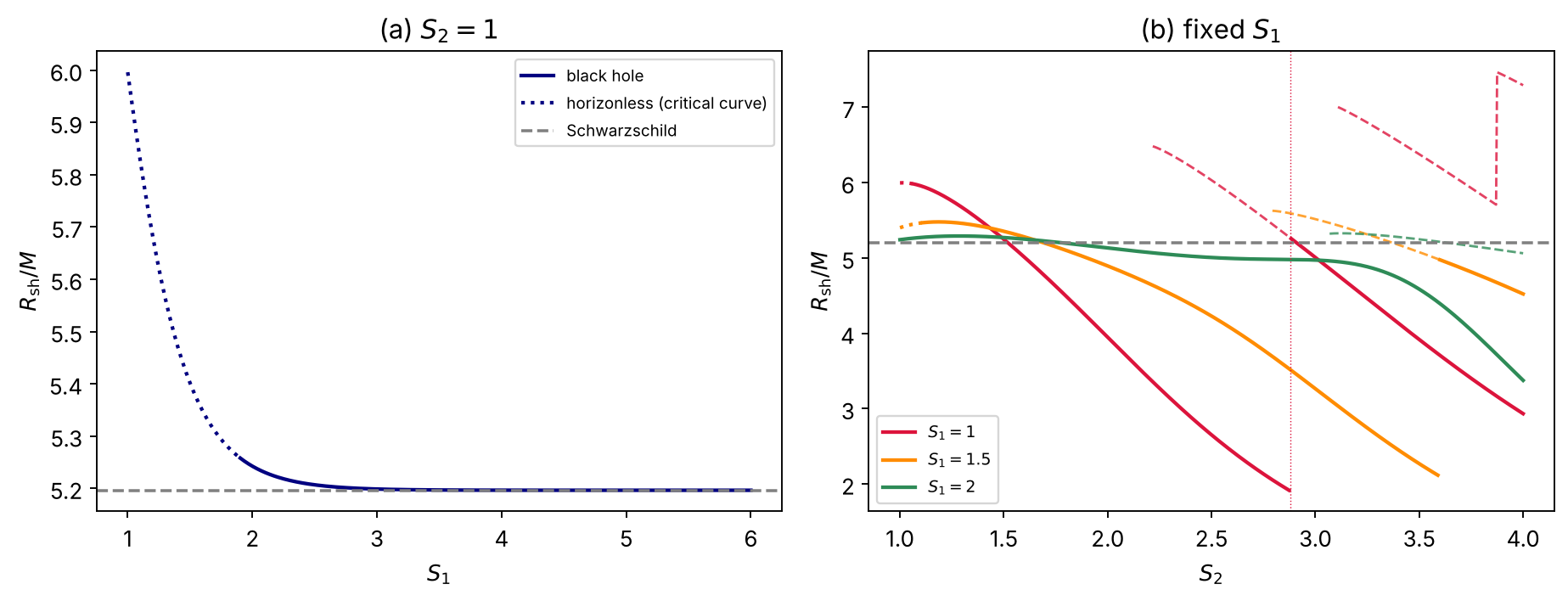}
\caption{Shadow radius $R_{\rm sh}$ of the non-rotating Lee-Wick black hole as a function of the coupling parameter $S_1$ (left panel, with $S_2=1$ fixed) and as a function of $S_2$ (right panel, for {$S_1=1$, $1.5$, and $2$}). The horizontal dashed line marks the Schwarzschild value $R_{\rm sh}^{\rm Schw}=3\sqrt{3}\,M$, to which the shadow radius converges as $S_1$ grows large. {Dotted segments: horizonless configurations, for which only a critical curve (no dark shadow) exists. Thin dashed curves in panel (b): critical impact parameter of the outer unstable photon sphere in configurations with two (or three) unstable photon spheres; this outer critical curve produces an additional bright photon ring outside the shadow (Fig.~\ref{fig:exotic}).} The thin dotted vertical line at $S_2\simeq2.879$ on the $S_1=1$ curve marks the horizon-bifurcation jump discussed in the text.}
\label{fig:shadow}
\end{figure*}

For a static observer at radial coordinate $r_o\to\infty$, the apparent angular radius of the shadow is $\sin^2\alpha_{\rm sh}=b_{\rm ph}^2 f(r_o)/r_o^2 \to b_{\rm ph}^2/r_o^2$, so that the linear shadow radius measured on the observer's celestial sphere is
\begin{equation}
R_{\rm sh} = b_{\rm ph} = \frac{r_{\rm ph}}{\sqrt{f(r_{\rm ph})}} .
\label{eq:Rsh}
\end{equation}
Using standard celestial coordinates $(\alpha,\beta)$ for a photon with impact parameters $(\xi,\eta)$ arriving at an observer at inclination $\theta_o$ \cite{Bardeen1973,Vincent2011}, the shadow boundary of the (non-rotating) Lee-Wick BH is the circle $\alpha^2+\beta^2=R_{\rm sh}^2$, since the spacetime is spherically symmetric. We quantify deviations from Schwarzschild using the fractional shadow deviation
\begin{equation}
\delta_{\rm sh} \equiv \frac{R_{\rm sh}(S_1,S_2)-R_{\rm sh}^{\rm Schw}}{R_{\rm sh}^{\rm Schw}} ,
\end{equation}
with $R_{\rm sh}^{\rm Schw}=3\sqrt3\,M\approx5.196\,M$.
For the rotating extension of the Lee-Wick BH \cite{SinghUpadhyayAli2022}, obtained via the Newman-Janis algorithm \cite{NewmanJanis1965,AzregAinou2014,DrakeSzekeres2000}, the shadow boundary is no longer circular and is more conveniently characterized by the Hioki-Maeda observables $(R_{\rm sh},\delta_{\rm s})$ \cite{HiokiMaeda2009} or by the endpoint asymmetry used in Refs.~\cite{KumarGhosh2020,BambiFreese2009}; we leave the fully spin-dependent shadow to future work and focus here on the non-rotating case, which already isolates the effect of $S_1,S_2$.

{\emph{Which critical curve bounds the shadow.} When $V_{\rm ph}(r)$ has more than one local maximum outside the horizon, a photon arriving from infinity with impact parameter $b$ is captured only if $1/b^2$ exceeds \emph{every} potential barrier along its path, i.e. if $b<\min_i b_{{\rm ph},i}$. Rays with $\min_i b_{{\rm ph},i}<b<\max_i b_{{\rm ph},i}$ cross the lower barrier, are reflected by the higher one, and escape. Hence the shadow edge is set by the \emph{global} maximum of $V_{\rm ph}$ (the smallest critical impact parameter), which we denote $R_{\rm sh}$; when this global maximum is the inner one, the outer unstable photon sphere is nevertheless visible as a second critical curve $b_{\rm out}>R_{\rm sh}$ outside the shadow. This criterion is applied consistently in Table~\ref{tab:shadow} and Figs.~\ref{fig:rph}--\ref{fig:deviation}.}

\subsection{Numerical results}

\begin{figure*}[tb]
\centering
\includegraphics[width=2\columnwidth]{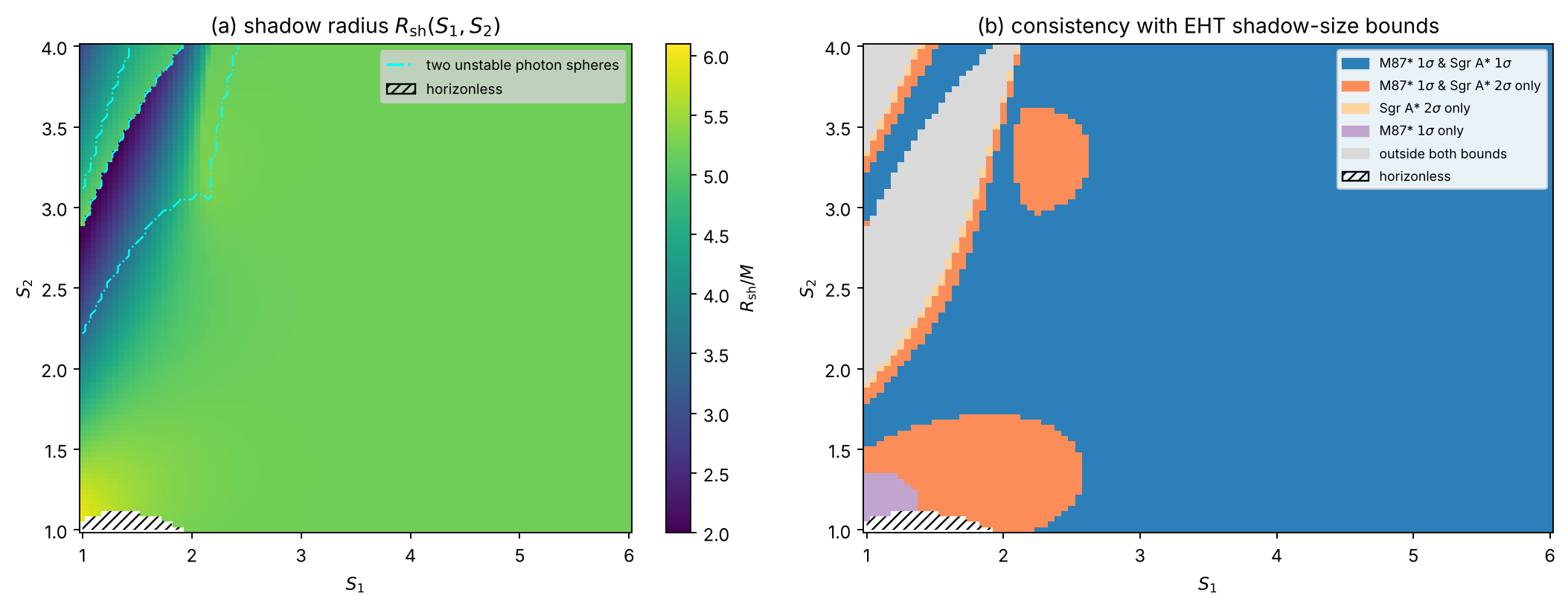}
\caption{{(a) Shadow radius $R_{\rm sh}(S_1,S_2)$ over the $(S_1,S_2)\in[1,6]\times[1,4]$ plane, evaluated at the global maximum of $V_{\rm ph}(r)$ outside the outer horizon (smallest critical impact parameter, see text). The hatched region is horizonless (no shadow), and the cyan dash-dotted contour encloses configurations with two unstable photon spheres. (b) Consistency of $R_{\rm sh}$ with the EHT shadow-size bounds: M87*, $4.31\lesssim R_{\rm sh}/M\lesssim6.08$ at $1\sigma$ \cite{EHT2019VI,Psaltis2020,Kocherlakota2021}; Sgr A*, $4.55\lesssim R_{\rm sh}/M\lesssim5.22$ ($1\sigma$) and $4.21\lesssim R_{\rm sh}/M\lesssim5.56$ ($2\sigma$), from the Keck- and VLTI-averaged EHT calibration \cite{EHT2022VI,Vagnozzi2023}. The Block-1 region $S_1\gtrsim2.65$ is consistent with all bounds for every $S_2\in[1,4]$, while two diagonal strong-coupling bands (gray) are excluded.}}
\label{fig:heatmap}
\end{figure*}

\begin{figure}[tb]
\centering
\includegraphics[width=\columnwidth]{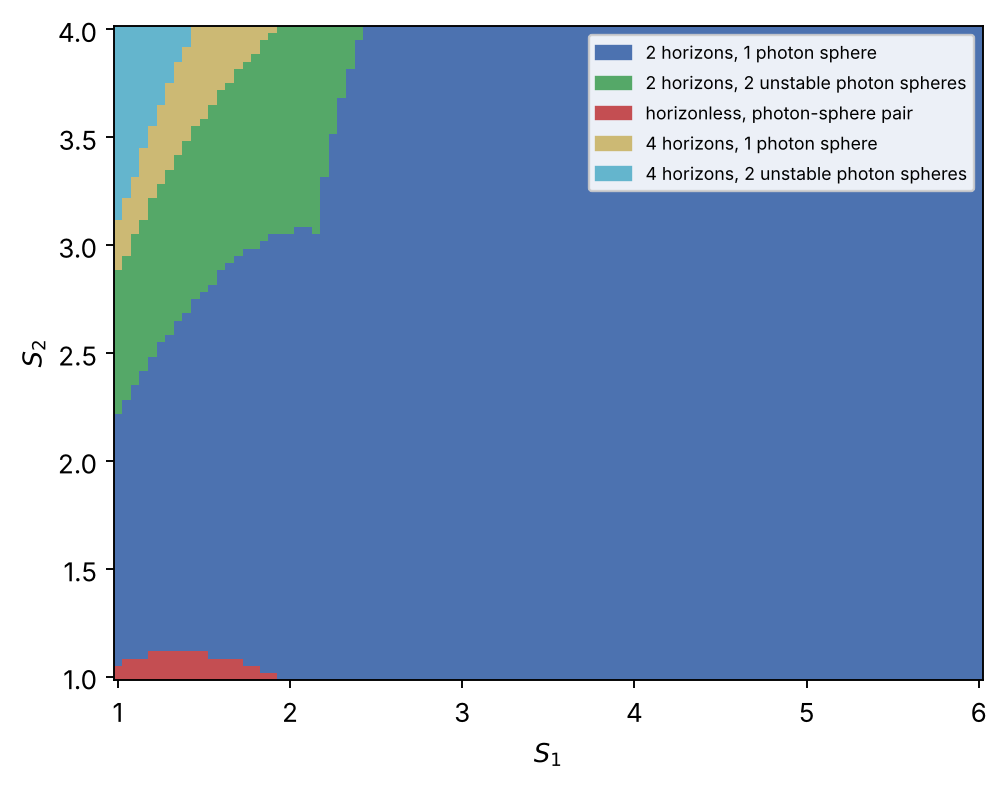}
\caption{{Classification of the Lee-Wick spacetime over the $(S_1,S_2)$ plane according to the number of horizons and of unstable photon spheres outside the outer horizon. Horizonless configurations (red) possess a pair of stable and unstable photon spheres and a regular core; the green and light-blue regions have two unstable photon spheres and therefore two critical curves; the boundary between two- and four-horizon regions is the horizon bifurcation discussed in the text.}}
\label{fig:class}
\end{figure}

\begin{figure*}[tb]
\centering
\includegraphics[width=2\columnwidth]{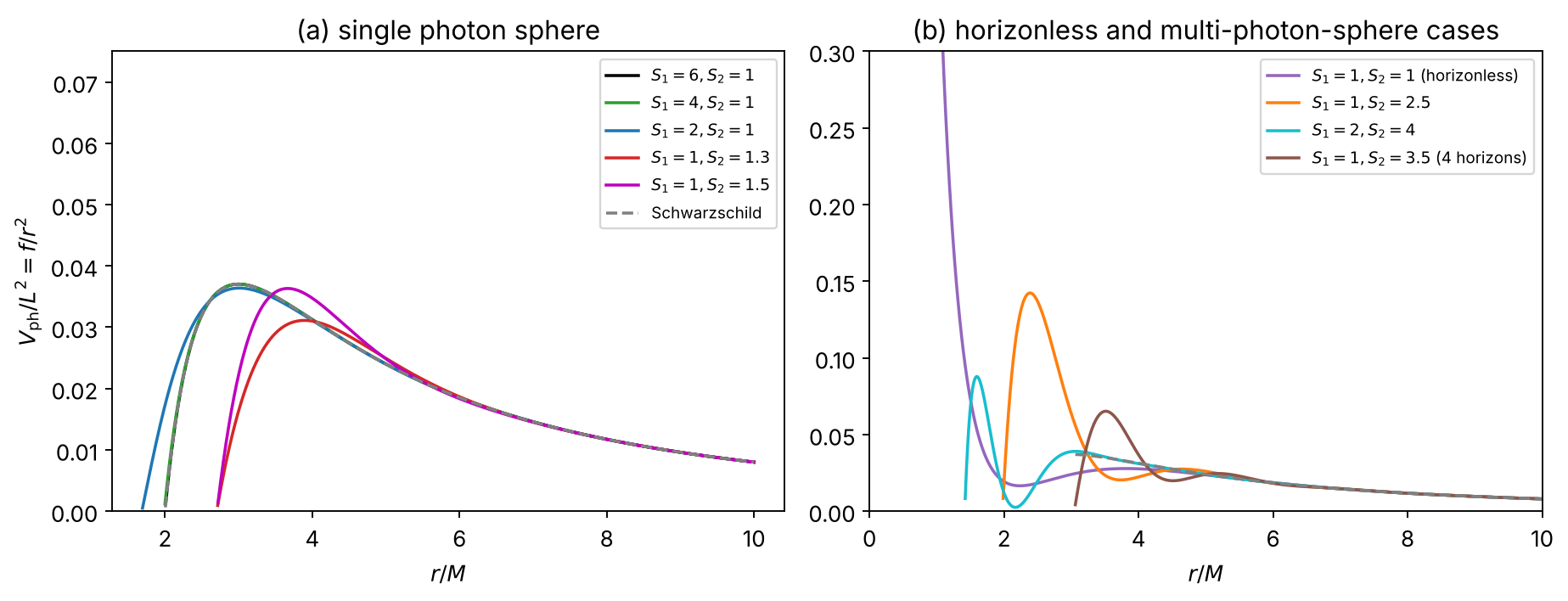}
\caption{Photon effective potential $V_{\rm ph}(r)/L^2=f(r)/r^2$ for several representative pairs of the coupling parameters $(S_1,S_2)$. {(a) Configurations with a single photon sphere,} spanning both the weak-coupling regime (black, green, and blue curves) and the strong-coupling regime (red and magenta curves), compared with the Schwarzschild potential (gray dashed curve, recovered in the limit of large $S_1$). {(b) A horizonless configuration (potential extending to $r\to0$, with a minimum corresponding to a stable photon sphere) and configurations with two unstable photon spheres, including a four-horizon case.} The radial location and height of each peak fix the photon sphere radius and shadow radius for the corresponding choice of $(S_1,S_2)$.}
\label{fig:potential}
\end{figure*}

\begin{figure}[tb]
\centering
\includegraphics[width=\columnwidth]{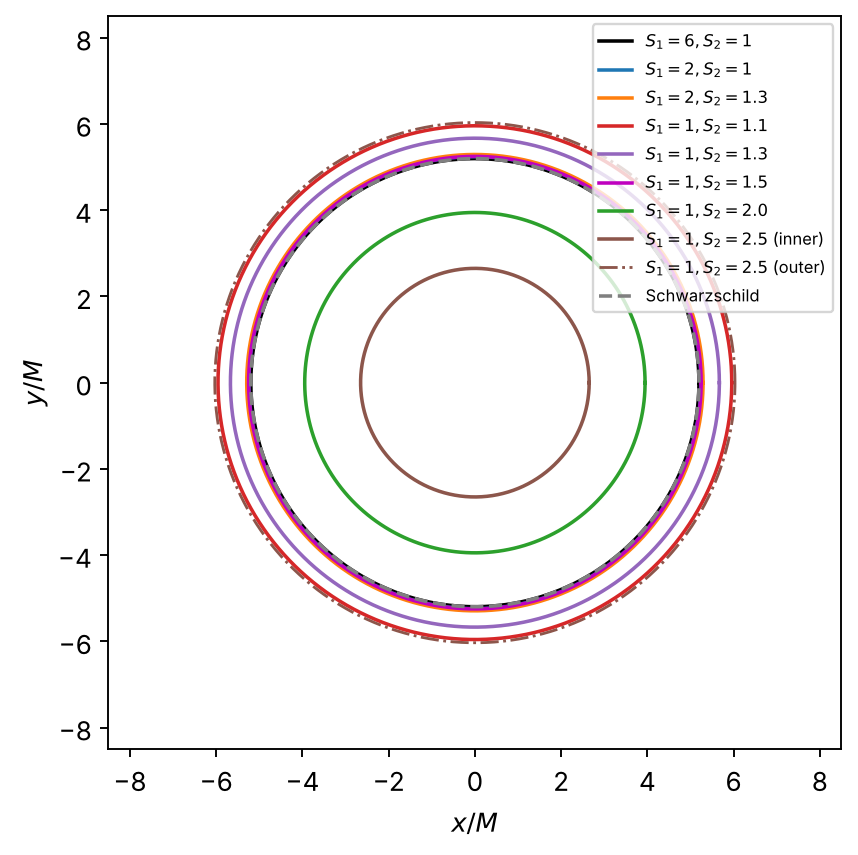}
\caption{Shadow silhouettes (celestial circles $\alpha^2+\beta^2=R_{\rm sh}^2$) of the non-rotating Lee-Wick black hole for {seven} representative parameter combinations, compared with the Schwarzschild silhouette (gray dashed circle). Because the non-rotating metric is spherically symmetric, each shadow boundary is exactly circular, with its radius set entirely by the value of $R_{\rm sh}$ for that choice of $S_1$ and $S_2$. {For $S_1=1$, $S_2=2.5$ both the inner (shadow-bounding) and the outer critical curves are shown.}}
\label{fig:silhouettes}
\end{figure}

We solved Eqs.~\eqref{eq:photonsphere}--\eqref{eq:Rsh} numerically {(bracketing of all sign changes of $f$ and of $2f-rf'$ on a dense radial grid, followed by Brent root refinement, and classification of every extremum of $V_{\rm ph}$)} over a broad grid of $(S_1,S_2)$ spanning both the Block-1 and Block-2 regimes identified in the hydrodynamical study \cite{DonmezMustafaYousafJavedSaidovAtamurotov2025}. Table~\ref{tab:shadow} lists the {number of horizons $N_h$,} outer horizon $r_+$, photon sphere radius $r_{\rm ph}$, shadow radius $R_{\rm sh}$, {Lyapunov exponent $\gamma$,} and the fractional deviation $\delta_{\rm sh}$ from Schwarzschild for {twenty-one} representative parameter combinations; Figs.~\ref{fig:shadow}--\ref{fig:silhouettes} present the same information graphically, from several complementary viewpoints.

\begin{table*}[t]
\caption{Number of horizons $N_h$, outer horizon radius $r_+$, radii of the unstable photon spheres $r_{\rm ph}$, shadow radius $R_{\rm sh}=b_{\rm ph}$ (all in units of $M$), fractional deviation $\delta_{\rm sh}=(R_{\rm sh}-R_{\rm sh}^{\rm Schw})/R_{\rm sh}^{\rm Schw}$, and Lyapunov exponent $\gamma$ [Eq.~\eqref{eq:lyap}] for representative combinations of $(S_1,S_2)$. {When two unstable photon spheres exist, both are listed; the shadow is bounded by the one with the smaller critical impact parameter, and the second value of $b$ (``outer ring'') gives the additional critical curve outside the shadow. For horizonless configurations no dark shadow forms, but a critical curve with impact parameter $b_c$ exists (the radius of the accompanying stable photon sphere is given in parentheses). Schwarzschild: $r_+=2$, $r_{\rm ph}=3$, $R_{\rm sh}=5.196$, $\gamma=\pi$.}}
\label{tab:shadow}
\begin{ruledtabular}
\begin{tabular}{cccccccl}
$S_1$ & $S_2$ & {$N_h$} & $r_+/M$ & $r_{\rm ph}/M$ & $R_{\rm sh}/M$ ($\delta_{\rm sh}$) & {$\gamma$} & {remarks} \\
\hline
1.0 & 1.0 & {0} & --- & {3.806 (stable: 2.237)} & {$b_c=5.996$} & {3.13} & {horizonless, regular core} \\
1.0 & 1.1 & {2} & 2.359 & 3.945 & 5.957 (+14.6\%) & {3.86} & \\
1.0 & 1.2 & {2} & 2.612 & 3.954 & 5.837 (+12.3\%) & {4.40} & \\
1.0 & 1.3 & {2} & 2.702 & 3.893 & 5.669 (+9.1\%) & {4.86} & \\
1.0 & 1.4 & {2} & 2.724 & 3.792 & 5.469 (+5.3\%) & {5.31} & \\
1.0 & 1.5 & {2} & 2.706 & 3.667 & 5.247 (+1.0\%) & {5.77} & \\
1.0 & 2.0 & {2} & 2.372 & 2.959 & 3.946 ($-$24.1\%) & {8.70} & \\
{1.0} & {2.5} & {2} & {1.977} & {2.387, 4.640} & {2.650 ($-$49.0\%)} & {12.8, 7.9} & {outer ring $b=6.032$} \\
{1.0} & {3.5} & {4} & {3.045} & {3.513, 5.177} & {3.916 ($-$24.6\%)} & {18.7, 11.7} & {outer ring $b=6.377$} \\
1.5 & 1.0 & {0} & --- & {3.108 (stable: 1.899)} & {$b_c=5.401$} & {2.55} & {horizonless, regular core} \\
1.5 & 1.5 & {2} & 2.330 & 3.364 & 5.354 (+3.0\%) & {4.12} & \\
1.5 & 4.0 & {4} & 2.583 & 3.082 & 4.522 ($-$13.0\%) & {10.96} & \\
2.0 & 1.0 & {2} & 1.686 & 3.008 & 5.242 (+0.9\%) & {2.85} & \\
2.0 & 1.3 & {2} & 2.038 & 3.144 & 5.291 (+1.8\%) & {3.34} & \\
2.0 & 2.0 & {2} & 2.160 & 3.008 & 5.132 ($-$1.2\%) & {3.78} & \\
{2.0} & {4.0} & {2} & {1.416} & {1.596, 3.065} & {3.374 ($-$35.1\%)} & {10.3, 5.7} & {outer ring $b=5.062$} \\
4.0 & 1.0 & {2} & 1.989 & 3.000 & 5.196 (0.0\%) & {3.14} & \\
5.0 & 1.0 & {2} & 1.998 & 3.000 & 5.196 (0.0\%) & {3.14} & \\
6.0 & 1.0 & {2} & 2.000 & 3.000 & 5.196 (0.0\%) & {3.14} & \\
6.0 & 3.0 & {2} & 2.000 & 3.000 & 5.196 (0.0\%) & {3.14} & \\
6.0 & 6.0 & {2} & 2.000 & 3.000 & 5.196 (0.0\%) & {3.14} & \\
\end{tabular}
\end{ruledtabular}
\end{table*}

\begin{figure}[tb]
\centering
\includegraphics[width=0.95\columnwidth]{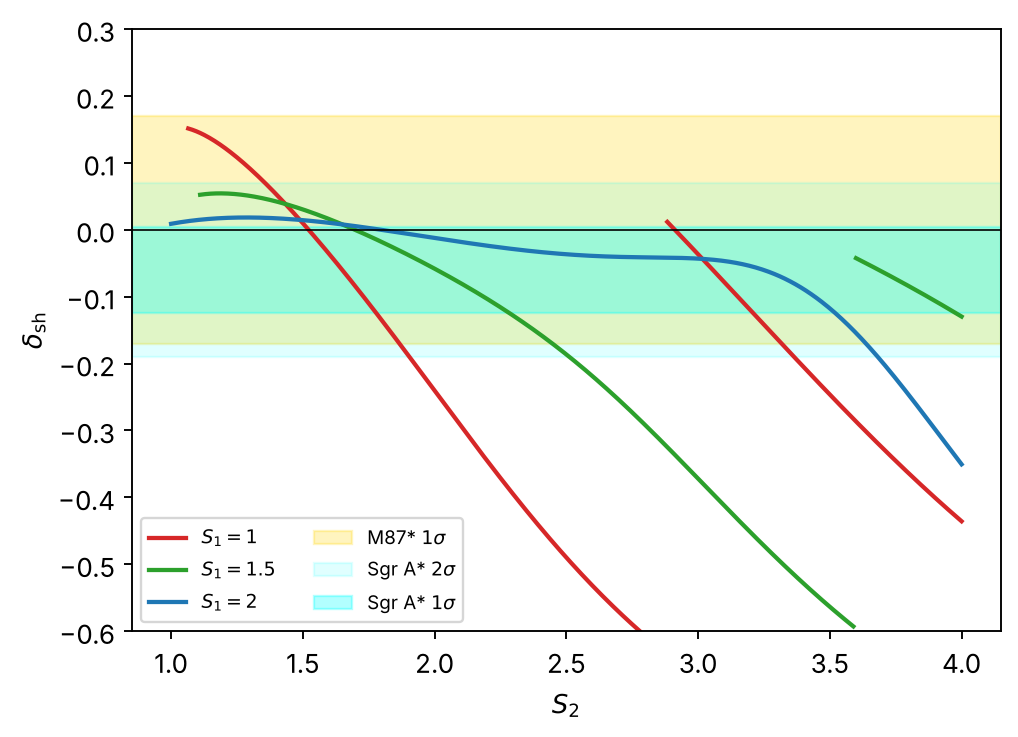}
\caption{Fractional shadow deviation $\delta_{\rm sh}=(R_{\rm sh}-R_{\rm sh}^{\rm Schw})/R_{\rm sh}^{\rm Schw}$ as a function of $S_2$, for $S_1=1$, $1.5$, and $2$, together with the {EHT bounds on the shadow size of M87* ($1\sigma$, gold) and Sgr A* ($1\sigma$ and $2\sigma$, cyan) quoted in Fig.~\ref{fig:heatmap}. Jumps correspond to the appearance of an inner, higher potential barrier or to the horizon bifurcation.} Portions of the curves lying outside the shaded bands correspond to parameter values that are disfavored by current horizon-scale imaging data.}
\label{fig:deviation}
\end{figure}

Several trends emerge from Table~\ref{tab:shadow} and Figs.~\ref{fig:shadow}--\ref{fig:deviation}. First, in the Block-1 regime ($S_2\lesssim2$, $S_1\gtrsim 2$), $R_{\rm sh}$ converges monotonically to the Schwarzschild value as $S_1$ increases (Fig.~\ref{fig:shadow}, left panel; Fig.~\ref{fig:heatmap}), mirroring the convergence of the ISCO and of the shock-cone opening angle reported in Ref.~\cite{DonmezMustafaYousafJavedSaidovAtamurotov2025}: the exponential suppression term dominates and washes out the oscillatory Lee-Wick correction, and already by $S_1\simeq4$ the shadow is indistinguishable from Schwarzschild to three significant figures. Second, for fixed small $S_1$ (Block-2, e.g. $S_1=1$), {the shadow is largest immediately after the horizon forms ($S_2\simeq1.07$, $R_{\rm sh}\simeq5.98\,M$, i.e. $\delta_{\rm sh}\simeq+15\%$) and then shrinks monotonically with increasing $S_2$, crossing the Schwarzschild value near $S_2\simeq1.52$ and reaching $\delta_{\rm sh}\simeq-24\%$ at $S_2=2$ (Table~\ref{tab:shadow}, Fig.~\ref{fig:deviation}). For $S_2\gtrsim2.2$ a second unstable photon sphere appears at larger radius (Fig.~\ref{fig:potential}b); since the inner barrier is the higher one, $R_{\rm sh}$ continues to decrease (to $\simeq2.0\,M$ at $S_2=2.8$), while an outer critical curve with $b_{\rm out}\simeq5.4$--$6.4\,M$ appears outside the shadow.} Third, for $S_1=1$ and $S_2=1$ the configuration is horizonless, consistent with Table~I of Ref.~\cite{DonmezMustafaYousafJavedSaidovAtamurotov2025}{; however, it is not devoid of photon spheres. It has an unstable photon sphere at $r=3.806\,M$ ($b_c=5.996\,M$) and a stable one at $r=2.237\,M$, and its core is regular ($f\to1$ as $r\to0$), so it is better described as a horizonless ultracompact object than as a naked singularity. Its critical curve exists, but rays with $b<b_c$ traverse the core and escape, so that no dark shadow is formed (Sec.~\ref{sec:class}, Fig.~\ref{fig:exotic})}. This non-monotonic behavior is the direct optical counterpart of the widened, asymmetric bow-shock morphology found for Block-2 accretion flows: a smaller/distorted photon sphere is consistent with the deeper, oscillatory effective potential wells identified via $V_{\rm eff}(r)$ in Ref.~\cite{DonmezMustafaYousafJavedSaidovAtamurotov2025}.

We note in addition a genuine, sharp discontinuity in the outer-horizon radius along the $S_1=1$ curve, visible in Figs.~\ref{fig:rph}(b) and \ref{fig:shadow}(b) (dotted vertical lines): at $S_2\simeq2.879$ a local minimum of $f(r)$ near $r\simeq3.2M$ touches zero and splits into a new pair of roots, so that $r_+$ jumps abruptly from $\simeq1.72M$ to $\simeq3.20M$ once this second, outer horizon becomes the largest root of $f(r)=0$. This is not a numerical artifact but a horizon-bifurcation of the type generic to multi-horizon Lee-Wick configurations \cite{Burzilla2023}{. At the bifurcation the two inner photon spheres are swallowed by the new outer horizon, so that the shadow radius jumps from $R_{\rm sh}\simeq1.9\,M$ to $\simeq5.2\,M$; the earlier outer critical curve then becomes the shadow boundary (Fig.~\ref{fig:shadow}b). An analogous bifurcation occurs at $S_2\simeq3.55$ for $S_1=1.5$. These jumps must be kept in mind when interpolating $R_{\rm sh}(S_1,S_2)$ across the Block-2 region.} {All curves in Figs.~\ref{fig:rph}--\ref{fig:deviation} are computed with the criterion of Sec.~III~B, which matters in the double-photon-sphere corner of the parameter plane.}

\subsection{{Horizon and photon-sphere structure, and implications for higher-order images}}
\label{sec:class}

{Figure~\ref{fig:class} summarizes the optical structure of the spacetime. Four qualitatively different cases occur in $[1,6]\times[1,4]$: (i) two horizons and a single unstable photon sphere (the generic, Schwarzschild-like case, which includes all of Block-1); (ii) two or four horizons and two unstable photon spheres separated by a stable one (a diagonal band at small $S_1$ and $S_2\gtrsim2.2$); (iii) the four-horizon region beyond the bifurcation; and (iv) horizonless configurations at $S_1\lesssim1.9$, $S_2\lesssim1.1$. In the horizonless case the configuration should not be regarded as a naked singularity: $f(r)$ remains positive everywhere and tends to unity at the center, the photon sphere pair always exists in this corner, and the configuration is an ultracompact, regular, horizonless object.}

{These structures have direct consequences for the image. (a) In case (ii) the shadow is bounded by the inner critical curve $R_{\rm sh}$, but the brightest ring of a geometrically thick, optically thin flow tracks the outer critical curve $b_{\rm out}$ (Fig.~\ref{fig:exotic}b), and lensed emission fills the annulus $R_{\rm sh}<\rho<b_{\rm out}$. A measurement of the ring diameter therefore constrains $b_{\rm out}$, not $R_{\rm sh}$, and the ring-to-shadow calibration that underlies EHT shadow inferences fails for such configurations. Similar double-ring structures were found for hairy black holes with two photon spheres \cite{Gan2021}. (b) In the horizonless case (iv), rays with $b<b_c$ pass through the regular core and escape, which produces a sequence of bright inner rings and removes the central brightness depression altogether (Fig.~\ref{fig:exotic}c,f); a resolved central depression, as observed for M87* and Sgr A* \cite{EHT2019I,EHT2022I}, thus disfavors this corner independently of any size argument. (c) The Lyapunov exponent of the shadow-bounding photon sphere, Eq.~\eqref{eq:lyap}, increases from $\gamma=\pi$ (Schwarzschild, and Block-1 for $S_1\gtrsim4$) to $\gamma\simeq3.9$--$5.8$ for $S_1=1$, $S_2=1.1$--$1.5$ and $\gamma\simeq8.7$ at $S_2=2$ (Table~\ref{tab:shadow}). The flux ratio of successive photon-ring sub-images, $e^{-\gamma}$, drops from $0.043$ (Schwarzschild) to $7.7\times10^{-3}$ for $(S_1,S_2)=(1,1.3)$ and $3.1\times10^{-3}$ for $(1,1.5)$; the $n\geq1$ sub-rings of Block-2 black holes are therefore much fainter and thinner, which is a distinctive target for future photon-ring interferometry \cite{Johnson2020}. Conversely, for $(S_1,S_2)=(2,1)$ one finds $\gamma\simeq2.85<\pi$, i.e. slightly brighter sub-rings than in Schwarzschild.}

\subsection{Comparison with EHT-scale constraints}
\label{sec:eht-shadow}

{The EHT has measured the angular diameter of the M87* emission ring to be $42\pm3\,\mu$as \cite{EHT2019VI} and that of Sgr A* to be $51.8\pm2.3\,\mu$as \cite{EHT2022I}. Translating these into bounds on the shadow size requires (i) independent mass-to-distance priors and (ii) a calibration factor between the emission-ring and critical-curve diameters, derived from GRMHD libraries of Kerr black holes \cite{EHT2019VI,EHT2022VI,Psaltis2020}. Following the standard procedure \cite{Psaltis2020,Kocherlakota2021,Vagnozzi2023}, we adopt for M87* the fractional deviation $\delta=-0.01\pm0.17$, i.e. $4.31\lesssim R_{\rm sh}/M\lesssim6.08$ at $1\sigma$, and for Sgr A* the Keck/VLTI-averaged bounds $4.55\lesssim R_{\rm sh}/M\lesssim5.22$ ($1\sigma$) and $4.21\lesssim R_{\rm sh}/M\lesssim5.56$ ($2\sigma$) \cite{EHT2022VI,Vagnozzi2023}.} We then require $R_{\rm sh}(S_1,S_2)$ to lie within these intervals.

{Overlaying these bounds on the full $(S_1,S_2)$ plane (Fig.~\ref{fig:heatmap}b) yields considerably more restrictive information than the three one-dimensional cuts of Fig.~\ref{fig:deviation}. (1) For $S_1\gtrsim2.65$ every $S_2\in[1,4]$ is consistent with both M87* and Sgr A* at $1\sigma$, and for $S_1\gtrsim2.15$ with M87* ($1\sigma$) and Sgr A* ($2\sigma$). (2) Two diagonal bands at small $S_1$ are excluded by both objects: the first follows the decrease of $R_{\rm sh}$ along the single-photon-sphere branch [e.g. $S_2\gtrsim1.87$ at $S_1=1$, $S_2\gtrsim2.45$ at $S_1=1.5$, $S_2\gtrsim3.63$ at $S_1=2$ for M87*], and the second lies just before the horizon bifurcation, inside the double-photon-sphere region. (3) Narrow re-entrant allowed strips exist just beyond the bifurcation (e.g. $2.88\lesssim S_2\lesssim3.31$ at $S_1=1$). (4) The $1\sigma$ upper bound of Sgr A*, $5.22\,M$, lies only $0.5\%$ above the Schwarzschild value, so that the mild shadow enlargement for $S_1\lesssim2.6$, $S_2\lesssim1.7$ (e.g. $S_2\lesssim1.52$ at $S_1=1$ and $S_2\lesssim1.72$ at $S_1=2$) is in tension at $1\sigma$ but allowed at $2\sigma$. Over the BH part of the scanned plane, $82\%$ of the area is consistent with both $1\sigma$ bounds and $7\%$ is inconsistent with both M87* ($1\sigma$) and Sgr A* ($2\sigma$).}

{These constraints should be interpreted with care. First, they treat $R_{\rm sh}$ as directly measured; our ray-traced images (Sec.~\ref{sec:numerics}) show that the ratio between the beam-convolved ring diameter and $2R_{\rm sh}$ varies from $\simeq1.10$ (Schwarzschild, Block-1) to $\simeq1.23$ for $(S_1,S_2)=(1,2)$, so that the Kerr-calibrated conversion can bias $R_{\rm sh}$ by up to $\sim10\%$ in the strong-coupling regime, and fails altogether in the double-photon-sphere and horizonless regions (Sec.~\ref{sec:class}). Second, spin is neglected; for the rotating Lee-Wick BH \cite{SinghUpadhyayAli2022} the shadow size depends mildly on spin and inclination, which would broaden the allowed bands. Third, the bounds inherit the systematic uncertainties of the mass and distance priors.} With these caveats, the shadow size provides a concrete, falsifiable target for horizon-scale tests of Lee-Wick gravity.

\section{Thick Accretion Flow Imaging and Polarization}
\label{sec:disk}

\begin{figure*}[tb]
\centering
\includegraphics[width=0.98\textwidth]{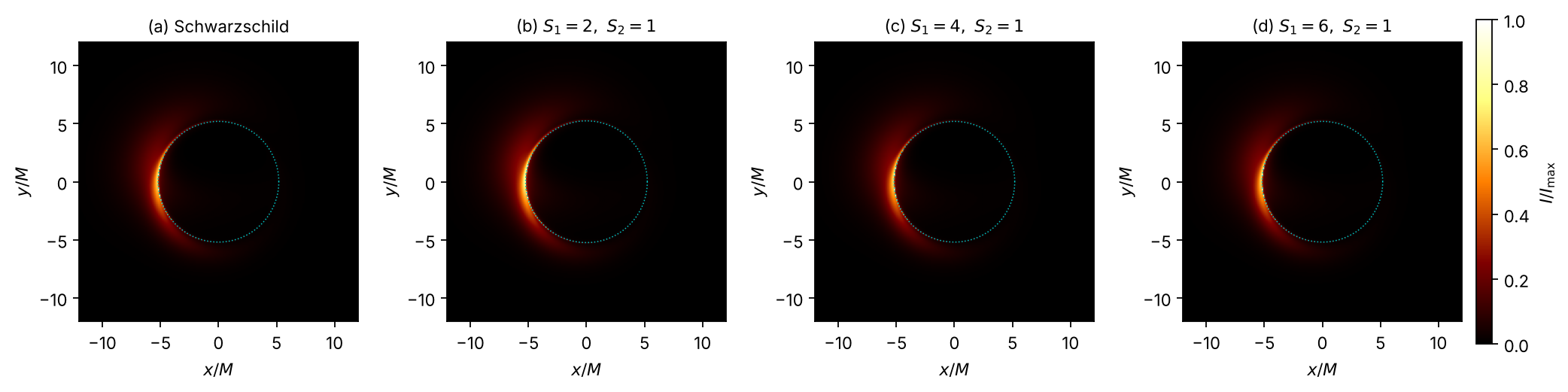}
\caption{Ray-traced 230~GHz images of the Lee-Wick black hole under thick-disk illumination for the weak-coupling sequence $S_1=2,4,6$ (with $S_2=1$ and observer inclination $\theta_o=70^\circ$ held fixed){, compared with Schwarzschild (panel a). Images are computed with the code and emission model of Sec.~\ref{sec:numerics} (thermal synchrotron, ballistic flow, toroidal field, $\sigma=0.3$, $240\times240$ pixels). The cyan dotted circles mark the critical curves $\alpha^2+\beta^2=R_{\rm sh}^2$.}}
\label{fig:img_S1}
\end{figure*}

\begin{figure*}[tb]
\centering
\includegraphics[width=0.98\textwidth]{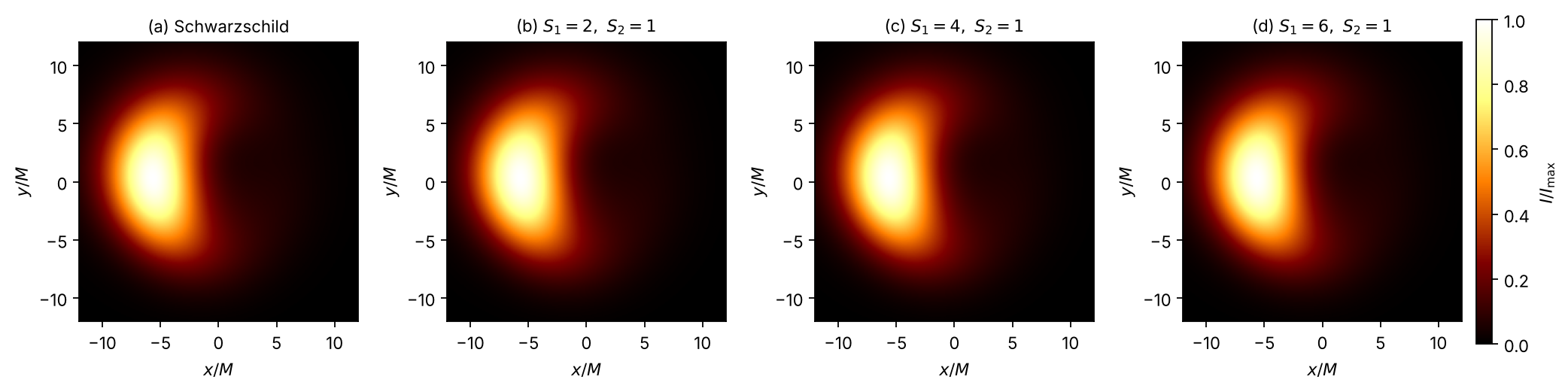}
\caption{The same weak-coupling sequence $S_1=2,4,6$ ($S_2=1$, $\theta_o=70^\circ$) {and Schwarzschild}, after convolving each ray-traced image with a {circular Gaussian beam of FWHM $20\,\mu$as, which corresponds to $5.3\,M$ for M87* ($\theta_g=GM/(c^2D)\simeq3.8\,\mu$as)} and mimics the angular resolution currently achievable with horizon-scale very-long-baseline interferometry. The convolution blends the primary photon ring with the fainter higher-order sub-images.}
\label{fig:img_S1_blur}
\end{figure*}

Motivated by the shock-cone/bow-shock structures obtained hydrodynamically in Ref.~\cite{DonmezMustafaYousafJavedSaidovAtamurotov2025}, and by the fact that EHT targets (M87*, Sgr A*) are surrounded by hot, optically thin, geometrically \emph{thick} flows rather than thin Novikov-Thorne disks \cite{NarayanYi1994,YuanNarayan2014}, we deliberately construct the imaging pipeline using two thick-disk prescriptions and do \emph{not} consider the thin-disk limit.

\subsection{Polarized radiative transfer}

Photon trajectories are obtained by backward ray tracing from the observer's image plane, integrating the null geodesic equations in the metric \eqref{eq:metric} (and its rotating extension \cite{SinghUpadhyayAli2022} when spin is included){; the specific implementation used in this work is described in Sec.~\ref{sec:numerics}}. Along each ray, the invariant polarized radiative-transfer equation for the Stokes vector $\mathbf{S}=(I,Q,U,V)/\nu^3$ is integrated,
\begin{equation}
\frac{d\mathbf{S}}{d\lambda} = \mathbb{J} - \mathbb{K}\,\mathbf{S},
\label{eq:rte}
\end{equation}
where $\lambda$ is an affine parameter, $\mathbb{J}=(j_I,j_Q,j_U,j_V)^T/\nu^2$ is the synchrotron emissivity vector, and $\mathbb{K}$ is the $4\times4$ Mueller matrix encoding absorption, Faraday rotation, and Faraday conversion \cite{Dexter2016,Moscibrodzka2018}. The polarization vector is parallel-transported along the null geodesic using the invariant Walker-Penrose constant {\cite{WalkerPenrose1970,ConnorsPiranStark1980}}, and the local emission/absorption coefficients are evaluated in the fluid rest frame and boosted to the observer frame via the local Doppler factor and the magnetic-field orientation on the sky.

The synchrotron emissivities depend on the electron number density $n_e(r,\theta)$, magnetic field $B(r,\theta)$, and (dimensionless) electron temperature $\Theta_e = k_BT_e/(m_ec^2)$, which are specified by the disk model.

\subsection{RIAF-like phenomenological model}

\begin{figure*}[tb]
\centering
\includegraphics[width=1.6\columnwidth]{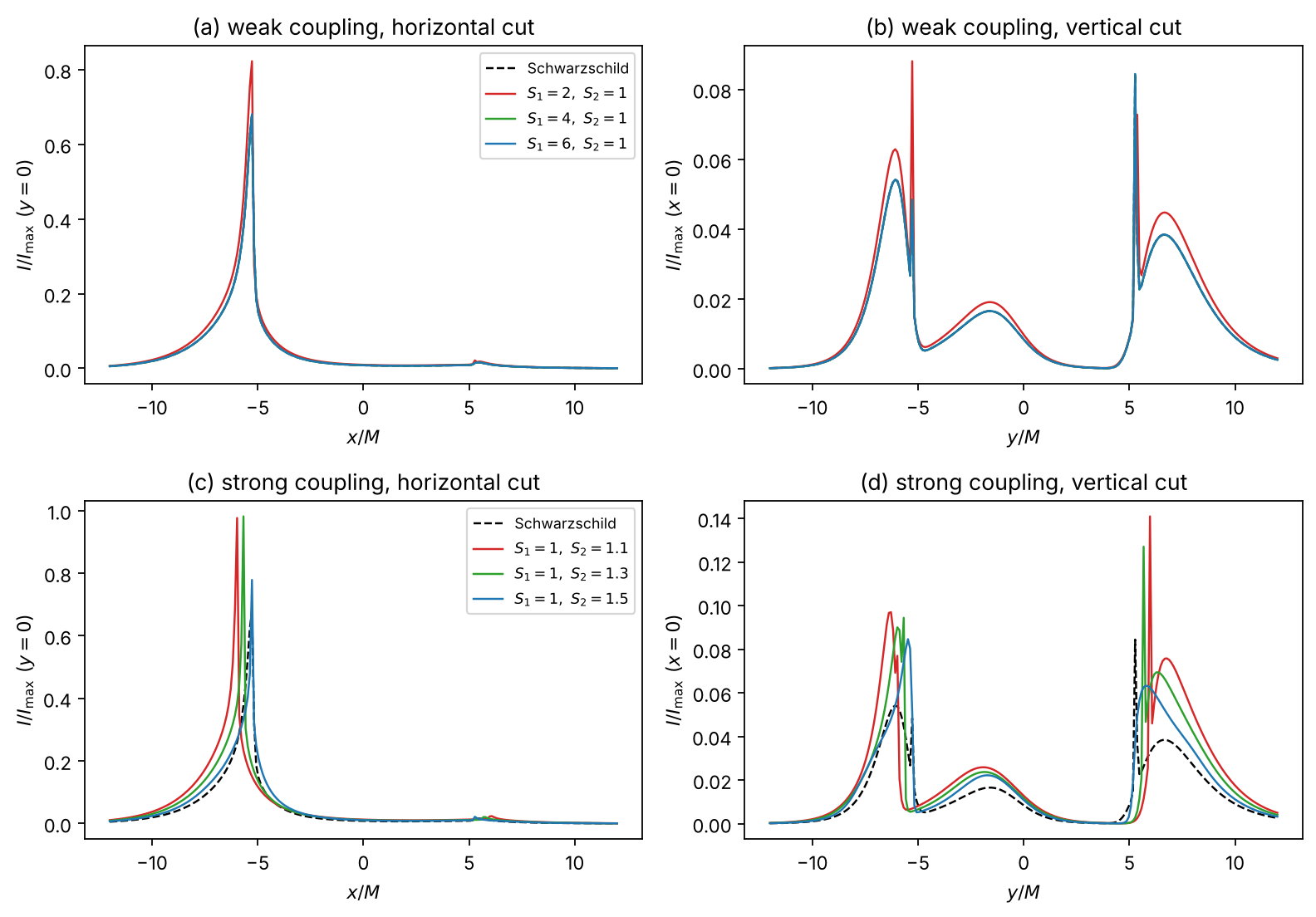}
\caption{Intensity cuts through the ray-traced images along the horizontal ($y=0$) and vertical ($x=0$) axes of the observer's screen, {for the weak-coupling sequence $S_1=2,4,6$ ($S_2=1$; panels a,b) and the strong-coupling sequence $S_2=1.1,1.3,1.5$ ($S_1=1$; panels c,d), at $\theta_o=70^\circ$, compared with Schwarzschild (dashed). The peaks mark the photon ring; in Block-1 they coincide with Schwarzschild to within one pixel, while in Block-2 they are displaced outward by up to $0.8\,M$, following $R_{\rm sh}$.}}
\label{fig:profiles}
\end{figure*}

\begin{figure*}[tb]
\centering
\includegraphics[width=0.98\textwidth]{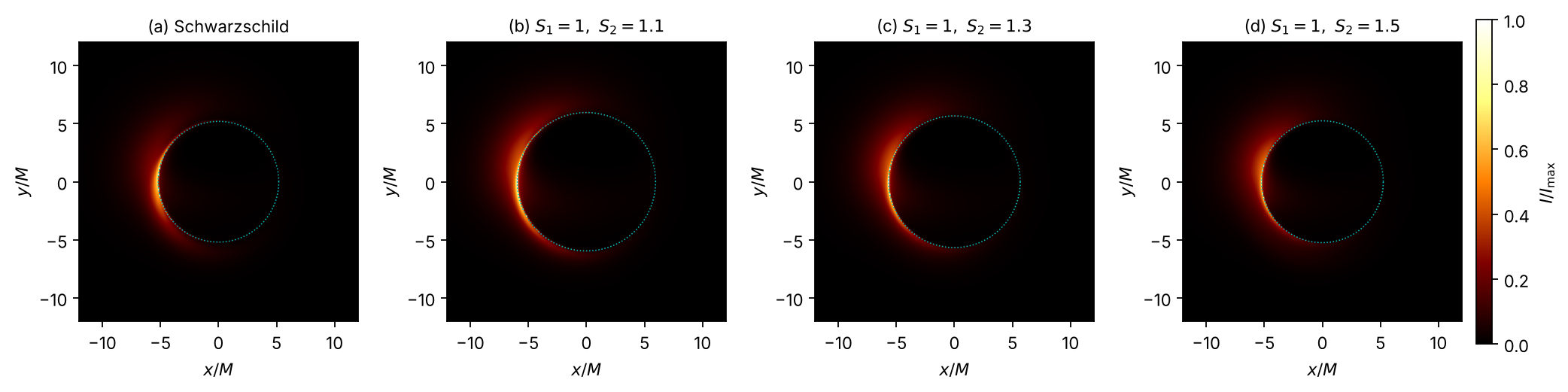}
\caption{Ray-traced {230~GHz} images of the Lee-Wick black hole under thick-disk illumination for the strong-coupling sequence $S_2=1.1,1.3,1.5$ (with $S_1=1$ and observer inclination $\theta_o=70^\circ$ held fixed){, compared with Schwarzschild. Emission model and markings as in Fig.~\ref{fig:img_S1}.} The bright ring visibly shrinks as $S_2$ increases, in direct correspondence with the shrinking shadow radius found analytically for this parameter sequence.}
\label{fig:img_S2}
\end{figure*}

Following the radiatively inefficient accretion flow (RIAF) framework \cite{NarayanYi1994,YuanNarayan2014,Broderick2011}, we adopt power-law radial profiles anchored at {a fixed reference radius $r_0=2M$ (rather than at $r_+$, so that the emission model is identical for all metrics and differences between images are purely due to the spacetime)},
\begin{align}
n_e(r) &= n_0 \left(\frac{r}{r_0}\right)^{-\beta_n}{e^{-(\theta-\pi/2)^2/2\sigma^2}}, \\
\Theta_e(r) &= \Theta_0 \left(\frac{r}{r_0}\right)^{-\beta_T}, \\
B(r) &= B_0 \left(\frac{r}{r_0}\right)^{-\beta_B},
\end{align}
{with $\beta_n=1.1$, $\beta_T=0.84$, $\beta_B=1$ \cite{Broderick2011}, $\Theta_0=50$, $B_0=50$\,G, and a Gaussian thickness envelope of angular half-width $\sigma$ (fiducial value $\sigma=0.3$ rad; we also show $\sigma=0.15$ and $0.6$ in Sec.~\ref{sec:numerics}).} The envelope keeps the emission concentrated around the equatorial region while remaining a genuinely thick, quasi-spherical flow{; since the flow is optically thin, $n_0$ only sets the overall normalization}. {The emissivity is the thermal-synchrotron fit of Ref.~\cite{Leung2011},
\begin{equation}
j_\nu=\frac{\sqrt2\pi e^2n_e\nu_s}{3K_2(1/\Theta_e)c}\left(X^{1/2}+2^{11/12}X^{1/6}\right)^2e^{-X^{1/3}},
\label{eq:jnu}
\end{equation}
with $X=\nu/\nu_s$, $\nu_s=\frac{2}{9}\nu_c\Theta_e^2\sin\theta_B$, $\nu_c=eB/(2\pi m_ec)$, and $\theta_B$ the angle between the photon and the magnetic field in the fluid frame. This is how the observing frequency enters: Eq.~\eqref{eq:jnu} is evaluated at the emitted frequency $\nu_{\rm em}=\nu_{\rm obs}/g$ with $\nu_{\rm obs}=230$~GHz, where $g$ is the redshift factor. With the above parameters $X\sim3$ at $r_0$, so that the 230~GHz emission is concentrated within $r\lesssim6M$, as in EHT-like models.} The fluid four-velocity is taken {to be the ballistic, marginally bound ``conical'' flow used in the analytic thick-disk models \cite{HouModel2023,ZhangThick2024}: $u_t=-1$, $u_\theta=0$, $u_\phi=\ell\sin^2\theta$, and $u^r=-\sqrt{1-f(1+\ell^2\sin^2\theta/r^2)}$, with specific angular momentum $\ell=3M$ (sub-critical in Schwarzschild, where capture requires $\ell<4M$). In the few regions where the radicand is negative (pockets with $f>1$ in strong coupling, or the core of horizonless configurations) we use a purely rotating flow with the same angular velocity ratio, capped at $v_{\hat\phi}=0.9$. This velocity field} remains well defined even where no circular orbits exist (relevant for the strong Block-2 regime, where the ISCO/photon sphere structure can be substantially altered).

\subsection{Analytic Hou disk model}

\begin{figure*}[tb]
\centering
\includegraphics[width=0.98\textwidth]{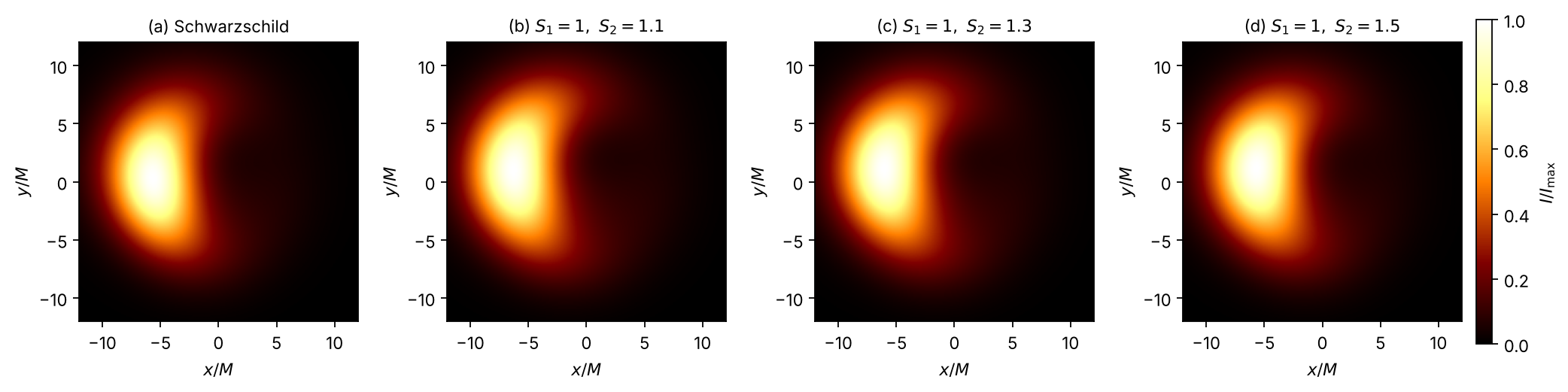}
\caption{The same strong-coupling sequence $S_2=1.1,1.3,1.5$ ($S_1=1$, $\theta_o=70^\circ$) {and Schwarzschild}, after convolving each ray-traced image with a Gaussian beam of {FWHM $20\,\mu$as ($5.3\,M$ for M87*)}, mimicking the angular resolution currently achievable with horizon-scale very-long-baseline interferometry.}
\label{fig:img_S2_blur}
\end{figure*}

\begin{figure*}[tb]
\centering
\includegraphics[width=0.85\textwidth]{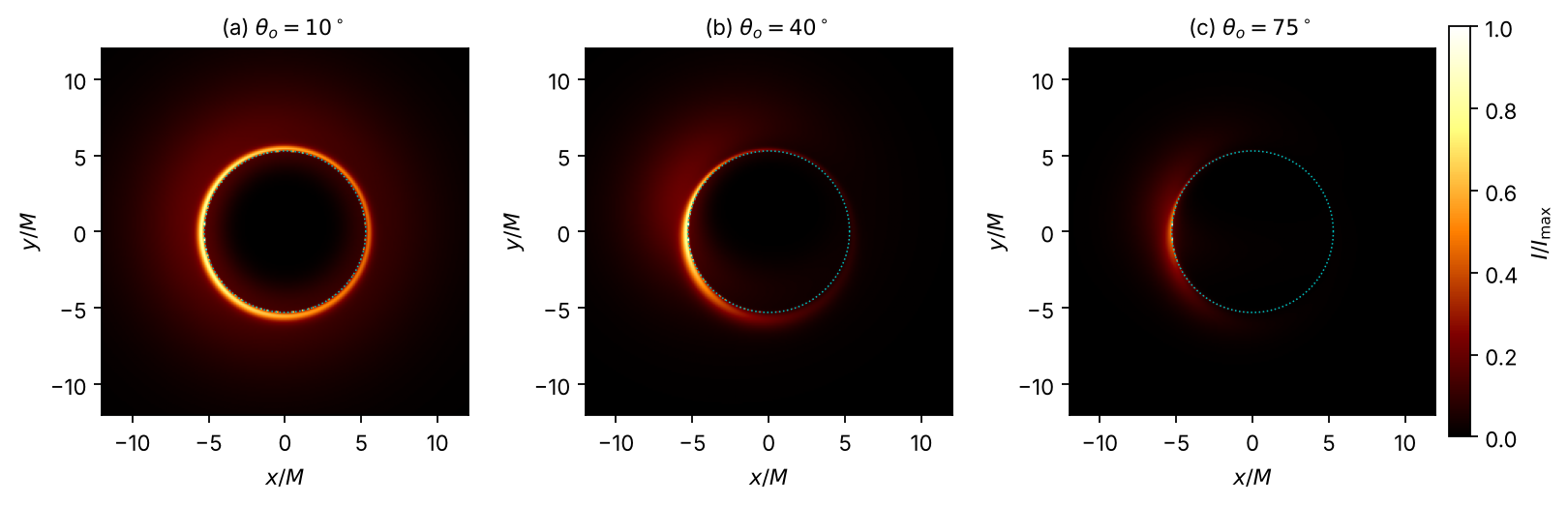}
\caption{Effect of the observer inclination $\theta_o=10^\circ,40^\circ,75^\circ$ at fixed $S_1=2,S_2=1.3$. Increasing $\theta_o$ leaves the ring size essentially unchanged but produces a pronounced Doppler-boosted crescent. {The images are free of polar-axis artifacts because each geodesic is integrated in its own orbital plane, which avoids the coordinate singularity of $(\theta,\phi)$ at the poles (Sec.~\ref{sec:numerics}).}}
\label{fig:img_incl}
\end{figure*}

As a complementary, more first-principles thick-disk prescription we {formulate} the analytic (``Hou'') disk model \cite{HouModel2023,ZhangThick2024}, in which the accreting fluid is confined to surfaces of constant polar angle, $u^\theta\equiv0$, so that the continuity equation reduces to
\begin{equation}
\frac{d}{dr}\left(\sqrt{-g}\,\rho\, u^r\right)=0
\;\;\Longrightarrow\;\;
\rho(r)=\rho_0\,\frac{\left(\sqrt{-g}\,u^r\right)\big|_{r_0}}{\sqrt{-g}\,u^r},
\label{eq:houcont}
\end{equation}
with $\rho_0=\rho(r_0)$ fixed at a reference radius (typically $r_0=r_+$). Projecting energy-momentum conservation along $u^\mu$ gives the first law
\begin{equation}
de = \frac{e+p}{\rho}\,d\rho ,
\end{equation}
where, defining the proton-to-electron temperature ratio $k=T_p/T_e$, the internal energy satisfies
\begin{equation}
e = \rho + \rho\,\frac{3}{2}(k+2)\frac{m_e}{m_p}\,\Theta_e .
\end{equation}
The fluid four-velocity $u^\mu$ is taken either as purely radial free fall or as a ballistic combination of radial infall and angular motion (the ballistic-approximation accretion flow, valid where gravity dominates the near-horizon dynamics) \cite{HouModel2023}; a purely circular flow is excluded because $u^\theta=0$ requires a nonzero radial velocity to avoid a divergent magnetic field. Equation~\eqref{eq:houcont} is evaluated using the Lee-Wick determinant $\sqrt{-g}=r^2\sin\theta$ (unchanged from Schwarzschild since the angular part of the metric \eqref{eq:metric} is unmodified) and the geodesic $u^r(r)$ appropriate to the Lee-Wick lapse function \eqref{eq:lapse}, so that all Lee-Wick corrections enter through $f(r)$ in the geodesic/radial-velocity profile rather than through the disk model itself. {In the present work the ballistic velocity field of this model is used in the numerical images (Sec.~\ref{sec:numerics}), but the thermodynamic profiles are taken from the RIAF-like prescription above; a full implementation of the Hou thermodynamics and magnetic field is left for future work.}

\subsection{Expected imaging signatures}

\begin{figure*}[tb]
\centering
\includegraphics[width=0.85\textwidth]{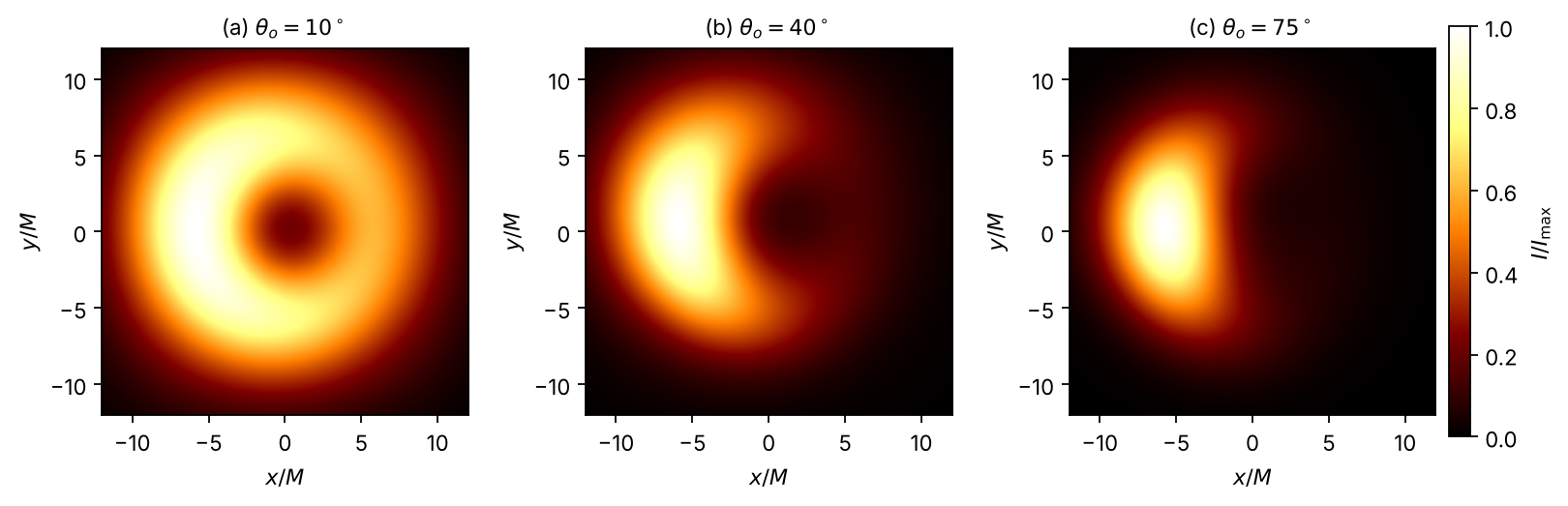}
\caption{The same inclination sequence $\theta_o=10^\circ,40^\circ,75^\circ$ (fixed $S_1=2$, $S_2=1.3$), after convolving each ray-traced image with a Gaussian beam of {FWHM $20\,\mu$as ($5.3\,M$ for M87*)}, mimicking the angular resolution currently achievable with horizon-scale very-long-baseline interferometry. The Doppler-boosted crescent at high inclination remains clearly visible even after blurring.}
\label{fig:img_incl_blur}
\end{figure*}

\begin{figure*}[tb]
\centering
\includegraphics[width=0.98\textwidth]{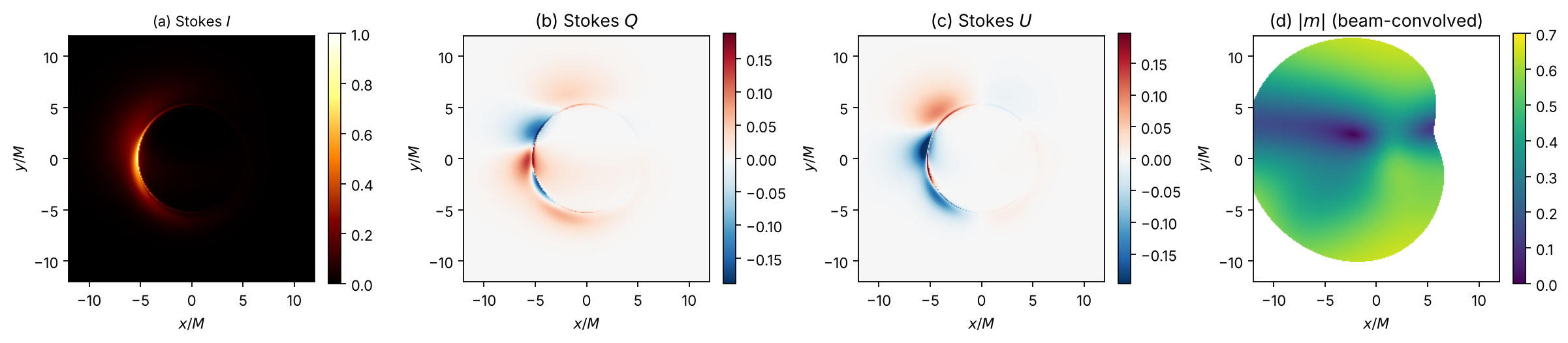}
\caption{Stokes parameters {$I$, $Q$, and $U$ (normalized to the peak of $I$) and beam-convolved fractional linear polarization $|m|=\sqrt{Q^2+U^2}/I$ for the fiducial thick-disk image ($S_1=2,S_2=1$, toroidal field, $\theta_o=70^\circ$). Stokes $V$ is not shown: in the present optically thin model without Faraday conversion and with intrinsic circular emissivity neglected, $V\equiv0$.}}
\label{fig:stokes}
\end{figure*}

Both disk models produce, at $230$~GHz, a central brightness depression (the shadow) surrounded by a bright, asymmetric photon ring whose diameter tracks $R_{\rm sh}(S_1,S_2)$ from Sec.~\ref{sec:shadow}, with higher-order (``$n=1,2,\dots$'') sub-images nested just inside the direct-emission ring, as is generic for RIAF-like and Hou-model thick disks around compact objects \cite{ZhangThick2024,GrallaHolzWald2019}. Guided by the hydrodynamic shock-cone results of Ref.~\cite{DonmezMustafaYousafJavedSaidovAtamurotov2025}, we expect:
\begin{itemize}
\item In the \textbf{Block-1} regime, images nearly indistinguishable from Schwarzschild: a circular, only mildly asymmetric photon ring, smooth linear-polarization ticks following the near-axisymmetric magnetic field threading the narrow, symmetric shock cone.
\item In the \textbf{Block-2} regime, a smaller {(or, close to the horizon-formation threshold, larger)} photon ring tracking $R_{\rm sh}$ of Fig.~\ref{fig:shadow}, with brightness asymmetries and depolarization patches correlated with the broad, asymmetric bow-shock/turbulent structure found in the GRHD simulations; the time-variable shock-cone dynamics (Figs.~12,17 of Ref.~\cite{DonmezMustafaYousafJavedSaidovAtamurotov2025}) should manifest as enhanced intra-night image variability, analogous to the flux and morphology variability reported by the EHT for Sgr A* \cite{EHT2022I}.
\end{itemize}
{These expectations are tested quantitatively in Sec.~\ref{sec:numerics}.}

\subsection{Numerical ray-traced images}
\label{sec:numerics}

\begin{figure*}[tb]
\centering
\includegraphics[width=0.98\textwidth]{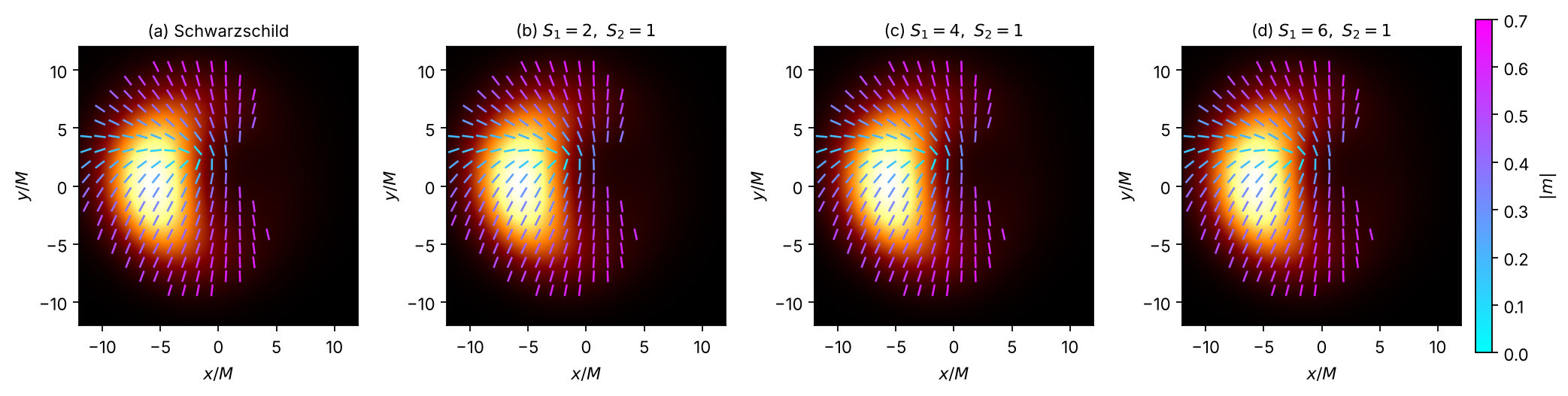}
\caption{Polarization images of the thick-disk emission for the weak-coupling sequence $S_1=2,4,6$ ($S_2=1$, $\theta_o=70^\circ$) {and Schwarzschild, for a toroidal field}, showing the {beam-convolved} total intensity map with superimposed electric-vector position-angle (EVPA) ticks, colored by the local degree of linear polarization $|m|$. {Ticks are shown where $I>0.08\,I_{\max}$.}}
\label{fig:pol_S1}
\end{figure*}

\begin{figure*}[tb]
\centering
\includegraphics[width=0.98\textwidth]{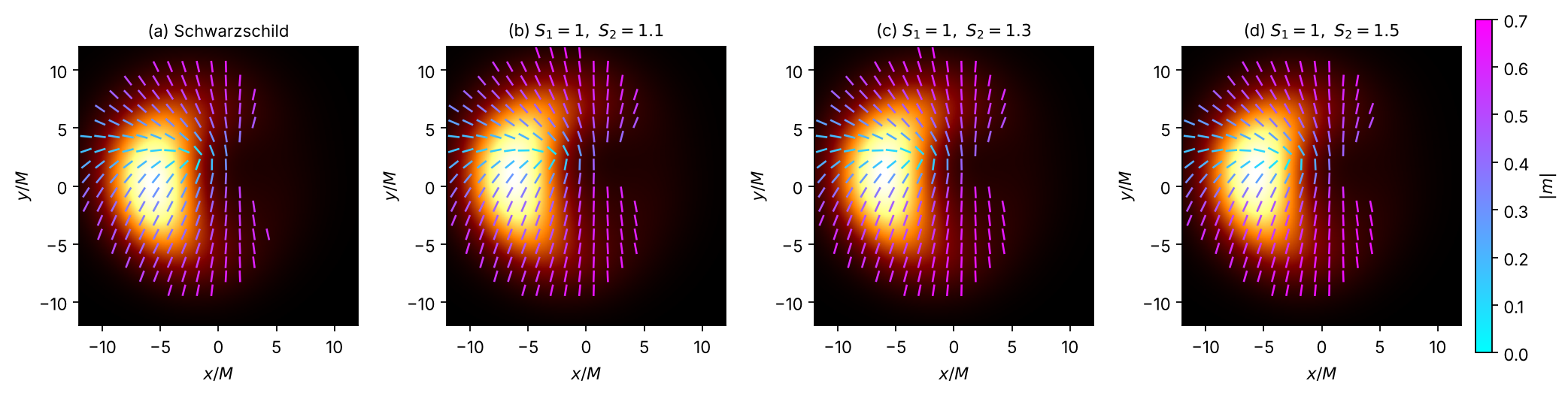}
\caption{Polarization images of the thick-disk emission for the strong-coupling sequence $S_2=1.1,1.3,1.5$ ($S_1=1$, $\theta_o=70^\circ$) {and Schwarzschild; same conventions as Fig.~\ref{fig:pol_S1}.} {The degree of linear polarization is suppressed in the bright crescent, where differently oriented EVPAs from the direct image and the photon ring are averaged within the beam, and recovers away from it.}}
\label{fig:pol_S2}
\end{figure*}

{\emph{Code and geodesic integration.} All images in this paper were produced with our own Python code, written for this work; no public GRRT package was used. Because the metric is spherically symmetric, every null geodesic lies in a plane through the origin. For each pixel $(x,y)$ of the observer's screen, with $b=\sqrt{x^2+y^2}$, we define the in-plane basis $\hat e_1=\hat n_o$ (the direction to the observer, at inclination $\theta_o$) and $\hat e_2=(x\hat e_R+y\hat e_U)/b$, where $\hat e_R,\hat e_U$ are the screen axes, and integrate
\begin{align}
&\frac{dr}{d\lambda}=fp_r,\qquad \frac{d\psi}{d\lambda}=\frac{b}{r^2},\nonumber\\
&\frac{dp_r}{d\lambda}=-\frac{1}{2}\left(f'p_r^2-\frac{2b^2}{r^3}+\frac{f'}{f^2}\right),
\label{eq:planar}
\end{align}
which follow from $2H=fp_r^2+b^2/r^2-1/f=0$ with $E=1$. The position on the sky of the BH is $\hat x=\cos\psi\,\hat e_1+\sin\psi\,\hat e_2$, from which $(\theta,\phi)$ follow without any coordinate singularity; in particular, rays passing over the poles (which occur for $x\simeq0$ at low inclination) are integrated without the pixel artifacts that a $(\theta,\phi)$ Hamiltonian integration produces there. Rays start at $r_o=500\,M$ with $p_r$ fixed by the null condition and are integrated backward with a fourth-order Runge-Kutta scheme and step $d\lambda=h\,r^2/(r+b)\times{\rm min}[1,\max(0.05,(r-r_+)/0.5M)]$, $h=0.015$, which resolves the winding near the photon sphere ($\lesssim0.015$ rad per step). Integration stops when $r<1.01\,r_+$ (capture) or when the ray recedes beyond $1.02\,r_o$; no ray reached the maximum of 9000 steps. $f$ and $f'$ are tabulated on a logarithmic grid of $4\times10^5$ points. The screen covers $[-12,12]M\times[-12,12]M$ with $240\times240$ pixels ($0.1\,M$ per pixel). As a check, the critical curve of every image coincides with the circle $\alpha^2+\beta^2=R_{\rm sh}^2$ of Sec.~\ref{sec:shadow} (cyan circles in Figs.~\ref{fig:img_S1}, \ref{fig:img_S2}, \ref{fig:img_incl}), and the polarization transport was verified against the flat-space limit to machine precision.}

{\emph{Radiative transfer.} Along each ray we integrate the invariant optically thin transfer equation, $I_\nu=\int g^2 j_\nu(\nu_{\rm obs}/g)\,d\lambda$, with $g=1/(-k_\mu u^\mu)$ computed exactly from the photon momentum and the ballistic four-velocity, and $j_\nu$ from Eq.~\eqref{eq:jnu} with the RIAF-like profiles of Sec.~IV~B. Emission is accumulated for $r_+<r<30\,M$. Absorption, Faraday rotation and conversion are neglected (optically thin limit), i.e. $\mathbb K=0$ in Eq.~\eqref{eq:rte}. For the polarization, the electric vector in the fluid frame is $f^\mu\propto\epsilon^{\mu\nu\rho\sigma}u_\nu k_\rho b_\sigma$ (perpendicular to the photon direction and to the field $b^\mu$), with intrinsic fractional polarization $p_0=0.7$. It is transported to the observer through the Walker-Penrose constants \cite{WalkerPenrose1970,ConnorsPiranStark1980}
\begin{equation}
\kappa_1=r\left(k^tf^r-k^rf^t\right),\qquad \kappa_2=r^3\sin\theta\left(k^\theta f^\phi-k^\phi f^\theta\right),
\end{equation}
which are conserved for \emph{any} lapse $f(r)$ in Eq.~\eqref{eq:metric} [$\kappa_2$ follows from the Killing-Yano tensor $r^3\sin\theta\,d\theta\wedge d\phi$ and $\kappa_1$ from the closed conformal Killing-Yano tensor $r\,dt\wedge dr$]. On the screen the electric vector is $\mathbf{E}\propto(\kappa_1x+\kappa_2y,\ \kappa_1y-\kappa_2x)$, and each emission element contributes $dQ+i\,dU=p_0\,g^2j_\nu e^{2i\chi}d\lambda$ with EVPA $\chi$ measured east of north. We consider three field geometries in the fluid frame: toroidal ($B^{\hat\phi}$), mixed radial-toroidal ($B^{\hat r}=B^{\hat\phi}$), and vertical (parallel to the spin axis); the field strength profile is the same in all cases.}

{\emph{Image-domain observables.} To compare models quantitatively we compute, on images convolved with a $20\,\mu$as beam: the ring diameter $d$ (twice the median, over 180 position angles, of the radius of peak brightness), the brightness asymmetry $A=(I_{\max}-I_{\min})/(I_{\max}+I_{\min})$ along the ring, the net linear polarization $|m|_{\rm net}=|\sum(Q+iU)|/\sum I$, and the rotationally symmetric polarization mode \cite{Palumbo2020}
\begin{equation}
\beta_2=\frac{\int\!\!\int (Q+iU)\,e^{-2i\varphi}\rho\,d\rho\,d\varphi}{\int\!\!\int I\,\rho\,d\rho\,d\varphi},
\end{equation}
over $\rho<10\,M$, with $\varphi$ the position angle east of north. A purely radial EVPA pattern has $\angle\beta_2=0^\circ$ and a purely azimuthal one $\angle\beta_2=180^\circ$.}

Figure~\ref{fig:img_S1} shows the resulting images for the Block-1 sequence $S_1=2,4,6$ ($S_2=1$, $\theta_o=70^\circ$): a dark shadow surrounded by a bright, asymmetric ring of higher-order images, with a pronounced Doppler-brightened crescent on the approaching side, essentially converged for $S_1\gtrsim4$ as anticipated from Fig.~\ref{fig:shadow}. {The sharp-image ring diameters are $d=10.58$, $10.61$, $10.58$, $10.58\,M$ for Schwarzschild and $S_1=2,4,6$, respectively, i.e. identical within the pixel size.} Figure~\ref{fig:img_S1_blur} shows the same sequence convolved with {the $20\,\mu$as beam}; the primary and higher-order images become blended. Figure~\ref{fig:profiles} shows the corresponding intensity cuts along the $x$- and $y$-axes{, which coincide with Schwarzschild in Block-1.}

Figure~\ref{fig:img_S2} and its blurred counterpart Fig.~\ref{fig:img_S2_blur} show the Block-2 sequence $S_2=1.1,1.3,1.5$ ($S_1=1$, $\theta_o=70^\circ$): here the ring visibly \emph{shrinks} as $S_2$ increases, in direct agreement with the $R_{\rm sh}(S_2)$ trend of Fig.~\ref{fig:shadow}(b) and Table~\ref{tab:shadow}{: the sharp-image diameters are $12.11$, $11.70$, and $10.78\,M$, i.e. $+14\%$, $+11\%$, and $+2\%$ relative to Schwarzschild, and the beam-convolved ones $11.29$, $10.72$, $10.07\,M$ versus $10.24\,M$}. Figure~\ref{fig:img_incl} and Fig.~\ref{fig:img_incl_blur} illustrate the effect of the observer inclination at fixed $S_1=2,S_2=1.3$: at $\theta_o=10^\circ$ the ring is nearly axisymmetric {($A=0.24$)}, while at $\theta_o=75^\circ$ pronounced Doppler beaming produces a bright crescent on one side and a faint, thin arc on the other {($A=0.85$)}, exactly the morphology anticipated qualitatively above and reported for Kerr and modified-gravity thick disks alike \cite{ZhangThick2024}.

Finally, Fig.~\ref{fig:stokes} shows the Stokes {$I$, $Q$, $U$ maps and the beam-convolved polarization fraction} for the fiducial case $S_1=2,S_2=1,\theta_o=70^\circ$, and Figs.~\ref{fig:pol_S1}--\ref{fig:pol_S2} show the polarization images (intensity with EVPA ticks colored by $|m|$) across the $S_1$ and $S_2$ sequences. {The fractional polarization is suppressed in the bright crescent, where the EVPAs of the direct and lensed emission differ and are averaged within the beam. This depolarization is not imposed but emerges from the transport itself.}

\begin{figure*}[tb]
\centering
\includegraphics[width=0.85\textwidth]{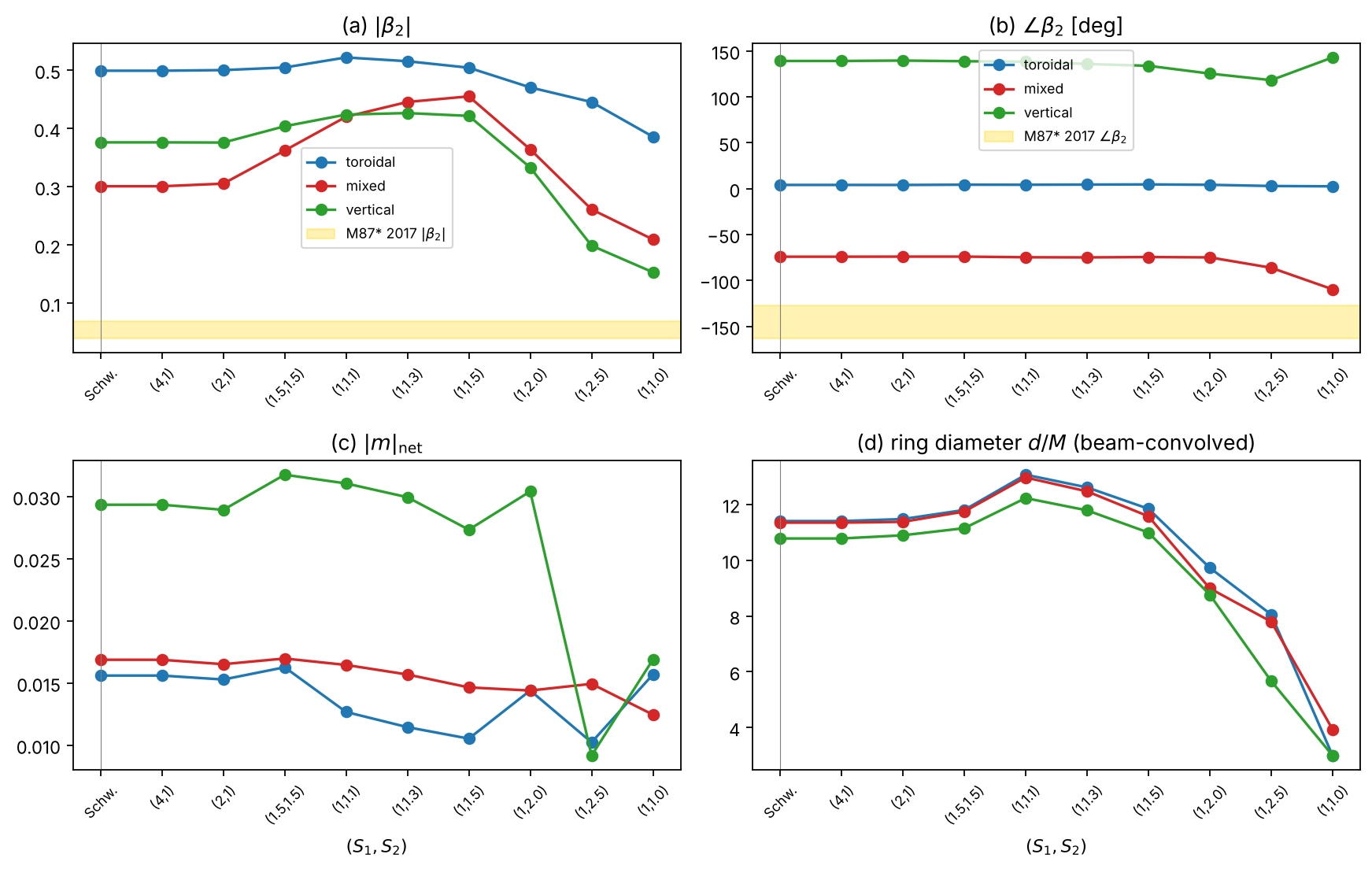}
\caption{{Quantitative comparison of beam-convolved ($20\,\mu$as) image observables between Schwarzschild and Lee-Wick black holes at the M87*-like inclination $\theta_o=17^\circ$, for toroidal (blue), mixed radial-toroidal (red), and vertical (green) magnetic fields: (a) amplitude and (b) phase of the $\beta_2$ coefficient, (c) net linear polarization, and (d) ring diameter. The last two configurations, $(1,2.5)$ and $(1,1)$, have two unstable photon spheres and no horizon, respectively. Gold bands: M87* 2017 ranges of $|\beta_2|$ and $\angle\beta_2$ \cite{EHT2021VII,EHT2021VIII}; our model, which has no Faraday rotation or turbulent field component, overpredicts $|\beta_2|$, so only the differences between metrics at fixed plasma model are meaningful.}}
\label{fig:polcomp}
\end{figure*}

{\emph{Quantitative comparison with Schwarzschild.} Figure~\ref{fig:polcomp} and Table~\ref{tab:polcomp} compare, at fixed plasma model, the image observables of Lee-Wick and Schwarzschild black holes at $\theta_o=17^\circ$. The main results are: (i) In Block-1 [$(4,1)$, $(2,1)$] all observables agree with Schwarzschild to $\lesssim1\%$ ($|\beta_2|$, $|m|_{\rm net}$) and $\lesssim0.1\,M$ in diameter, so that Block-1 is observationally indistinguishable at present. (ii) In Block-2 the beam-convolved ring diameter (toroidal field) changes by $+14\%$, $+11\%$, $+4\%$, and $-15\%$ for $S_2=1.1$, $1.3$, $1.5$, and $2.0$ ($S_1=1$). (iii) The $\beta_2$ \emph{phase} is remarkably robust: for all configurations with a single photon sphere it changes by $\lesssim1^\circ$ for the toroidal and mixed fields and by $\lesssim14^\circ$ for the vertical field, i.e. at low inclination $\angle\beta_2$ traces the magnetic-field geometry rather than the metric. (iv) The $\beta_2$ \emph{amplitude} is the most sensitive polarimetric discriminant: for the mixed field it increases from $0.301$ (Schwarzschild) to $0.420$, $0.446$, and $0.455$ ($+40\%$ to $+51\%$) for $S_2=1.1$, $1.3$, $1.5$ at $S_1=1$, and by $+20\%$ for $(1.5,1.5)$. We attribute this mainly to the larger Lyapunov exponent of Block-2 photon spheres (Sec.~\ref{sec:class}): the $n\geq1$ sub-images, whose EVPA pattern is rotated with respect to the direct image \cite{Himwich2020}, are fainter, so that the beam-averaged polarization pattern is less scrambled. (v) In configurations with two unstable photon spheres and in horizonless configurations, $\angle\beta_2$ rotates by $12^\circ$ and $36^\circ$ (mixed field), respectively, and $|\beta_2|$ decreases by $13\%$ and $30\%$, because additional rings with differently oriented EVPAs contribute within the beam. (vi) At high inclination ($\theta_o=70^\circ$) the differences are larger in phase: for the mixed field $\angle\beta_2$ changes from $-168^\circ$ (Schwarzschild) to $-150^\circ$, $-147^\circ$, $-147^\circ$ for $S_2=1.1$, $1.3$, $1.5$ ($S_1=1$), and $|m|_{\rm net}$ for the toroidal field increases from $0.20$ to $0.25$--$0.26$. These are the distinctive polarization signatures of the Lee-Wick geometry within the present model; they are differential signatures, comparable in size to the effects of changing the field geometry, so that a test requires a joint fit of plasma and metric parameters.}

\begin{table}[tb]
\caption{{Beam-convolved image observables at $\theta_o=17^\circ$ for the mixed radial-toroidal field (toroidal-field values of $|\beta_2|$ in parentheses). $d$ is the ring diameter in units of $M$; $\Delta$ denotes the relative change with respect to Schwarzschild.}}
\label{tab:polcomp}
\begin{ruledtabular}
\begin{tabular}{lcccc}
{$(S_1,S_2)$} & {$d/M$} & {$|\beta_2|$} & {$\angle\beta_2$} & {$\Delta|\beta_2|$} \\
\hline
{Schw.} & {11.34} & {0.301 (0.499)} & {$-74^\circ$} & {---} \\
{(4,1)} & {11.34} & {0.301 (0.499)} & {$-74^\circ$} & {0\%} \\
{(2,1)} & {11.37} & {0.305 (0.500)} & {$-74^\circ$} & {+1\%} \\
{(1.5,1.5)} & {11.74} & {0.362 (0.505)} & {$-74^\circ$} & {+20\%} \\
{(1,1.1)} & {12.96} & {0.420 (0.522)} & {$-75^\circ$} & {+40\%} \\
{(1,1.3)} & {12.46} & {0.446 (0.516)} & {$-75^\circ$} & {+48\%} \\
{(1,1.5)} & {11.57} & {0.455 (0.504)} & {$-75^\circ$} & {+51\%} \\
{(1,2.0)} & {8.99} & {0.364 (0.471)} & {$-75^\circ$} & {+21\%} \\
{(1,2.5)} & {7.80} & {0.261 (0.445)} & {$-86^\circ$} & {$-13\%$} \\
{(1,1.0)} & {3.94} & {0.210 (0.386)} & {$-110^\circ$} & {$-30\%$} \\
\end{tabular}
\end{ruledtabular}
\end{table}

\begin{figure*}[tb]
\centering
\includegraphics[width=0.85\textwidth]{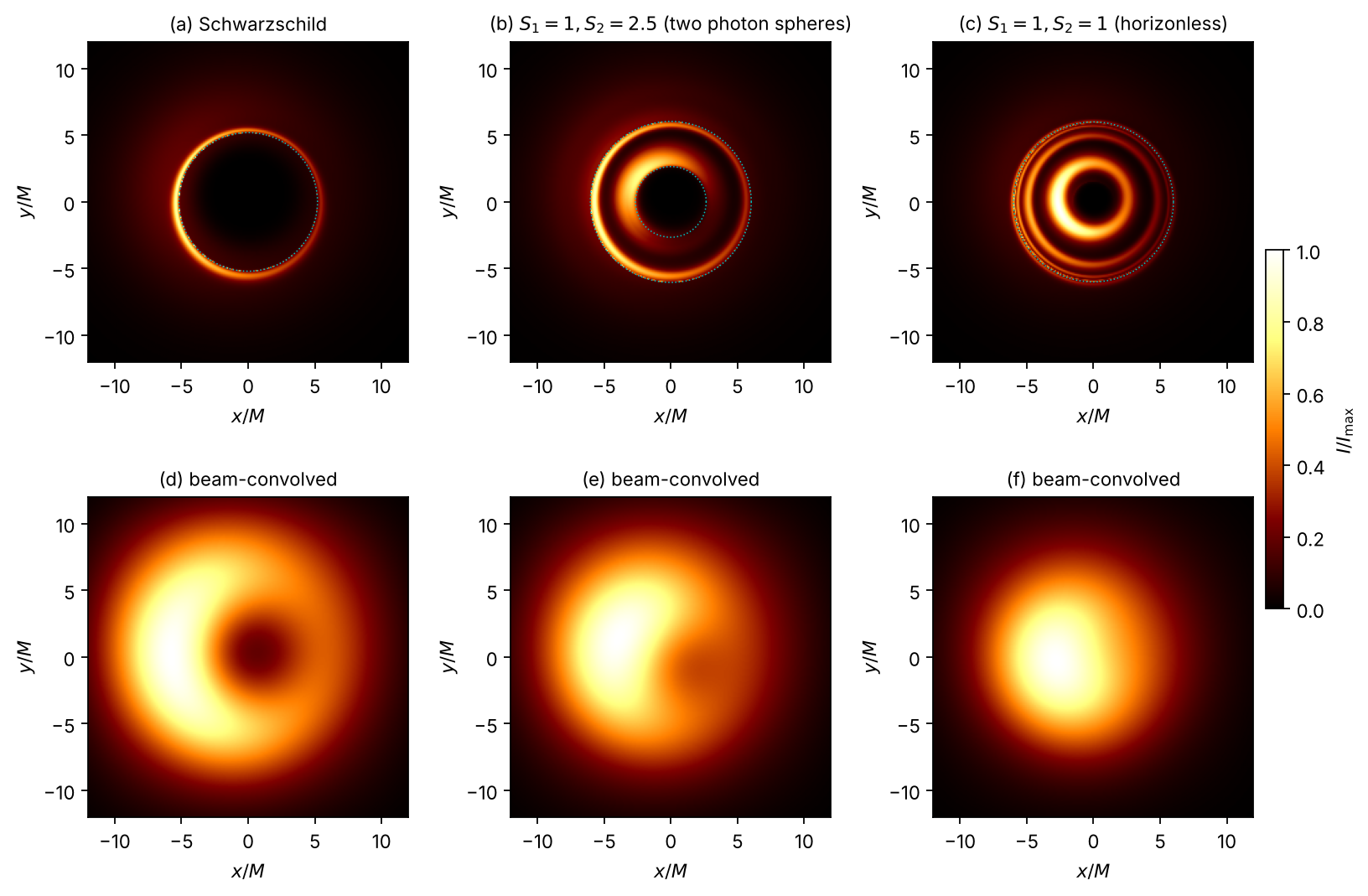}
\caption{{230~GHz images at $\theta_o=17^\circ$ (toroidal field) of (a) Schwarzschild, (b) a configuration with two unstable photon spheres ($S_1=1$, $S_2=2.5$; inner critical curve $R_{\rm sh}=2.65\,M$, outer one $b_{\rm out}=6.03\,M$), and (c) a horizonless configuration ($S_1=S_2=1$, $b_c=6.00\,M$). Cyan dotted circles: critical curves. Panels (d)--(f): the same images convolved with a $20\,\mu$as beam. In (b) the dark region is bounded by the inner critical curve while the bright ring follows the outer one; in (c) rays crossing the regular core produce nested inner rings and no central brightness depression survives the convolution.}}
\label{fig:exotic}
\end{figure*}

{\emph{Multiple photon spheres and horizonless configurations.} Figure~\ref{fig:exotic} illustrates the discussion of Sec.~\ref{sec:class}. For $(S_1,S_2)=(1,2.5)$, the dark central region has radius $\simeq R_{\rm sh}=2.65\,M$, while the brightest ring lies on the outer critical curve $b_{\rm out}=6.03\,M$, and the annulus between them is filled with lensed emission and a second, inner photon ring. After convolution, the image shows a ring with a strongly reduced central depression. For the horizonless configuration $(1,1)$ the image consists of the outer photon ring at $b_c$ and a set of bright inner rings formed by rays that cross the core; the beam-convolved image is centrally peaked. Both cases are therefore distinguishable from a Schwarzschild-like BH by morphology alone.}

\begin{figure*}[tb]
\centering
\includegraphics[width=0.98\textwidth]{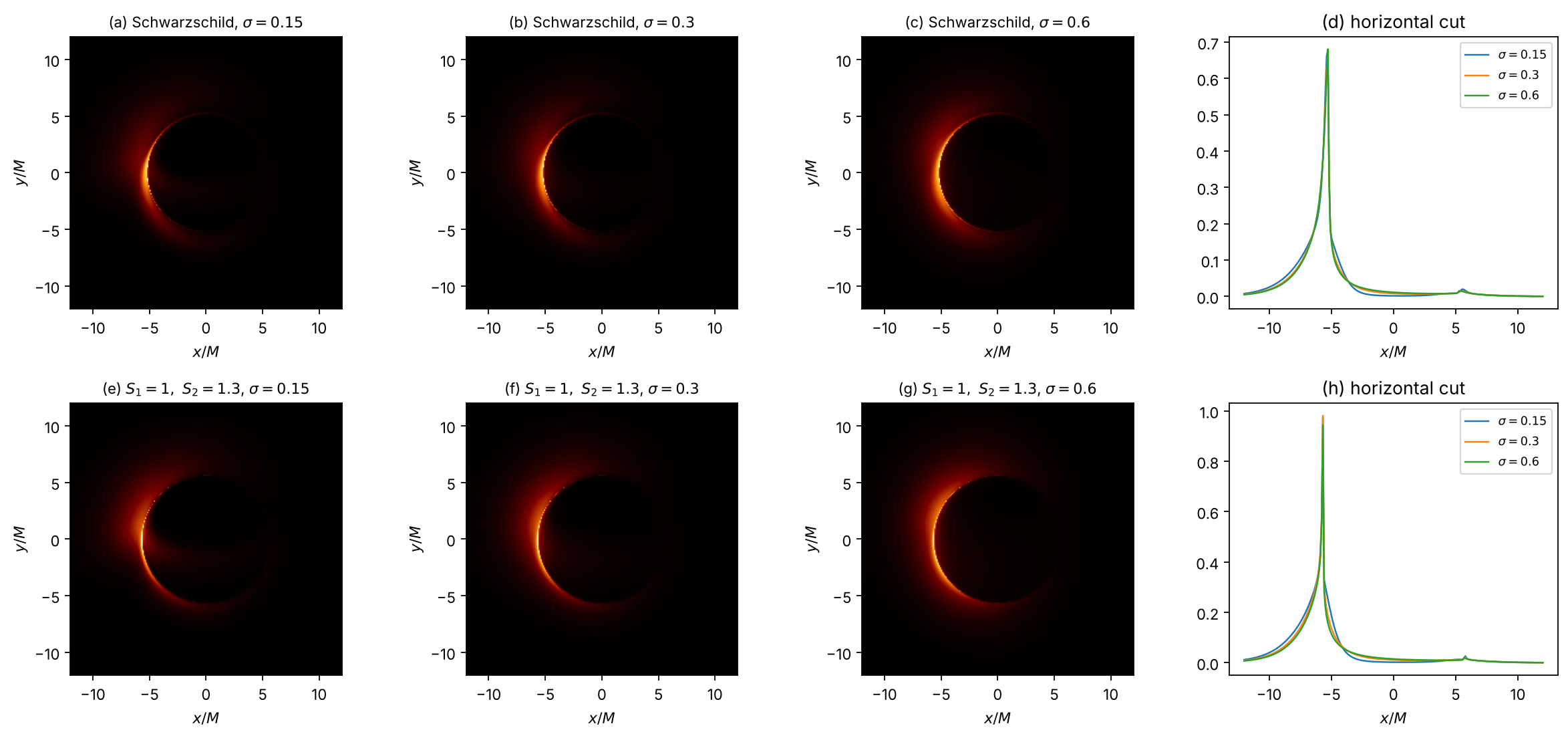}
\caption{{Dependence of the 230~GHz images on the disk thickness parameter $\sigma$ ($0.15$, $0.3$, $0.6$ rad) for Schwarzschild (top) and $S_1=1$, $S_2=1.3$ (bottom), at $\theta_o=70^\circ$ (toroidal field). Panels (d) and (h) show horizontal intensity cuts normalized to the maximum of each image.}}
\label{fig:sigma}
\end{figure*}

{\emph{Dependence on the disk thickness.} The fiducial thickness parameter is $\sigma=0.3$ rad. Figure~\ref{fig:sigma} compares $\sigma=0.15$, $0.3$, and $0.6$. The position of the sharp photon ring is unaffected ($d=10.58$ and $10.55\,M$ for Schwarzschild, $11.83$--$11.51\,M$ for $(1,1.3)$), as expected since it is fixed by the critical curve. In contrast, the total flux increases by a factor $\simeq3.1$ from $\sigma=0.15$ to $0.6$, the brightness asymmetry decreases from $0.86$ to $0.81$--$0.82$, and the beam-convolved ring diameter increases from $9.84$ to $10.89\,M$ (Schwarzschild) and from $9.85$ to $11.54\,M$ $(1,1.3)$, i.e. by $\sim10$--$17\%$. A change of disk thickness can therefore mimic in the blurred image a shift comparable to the Block-2 deviations, which is another reason why the shadow size should be inferred jointly with the flow geometry.}

\begin{figure*}[tb]
\centering
\includegraphics[width=0.98\textwidth]{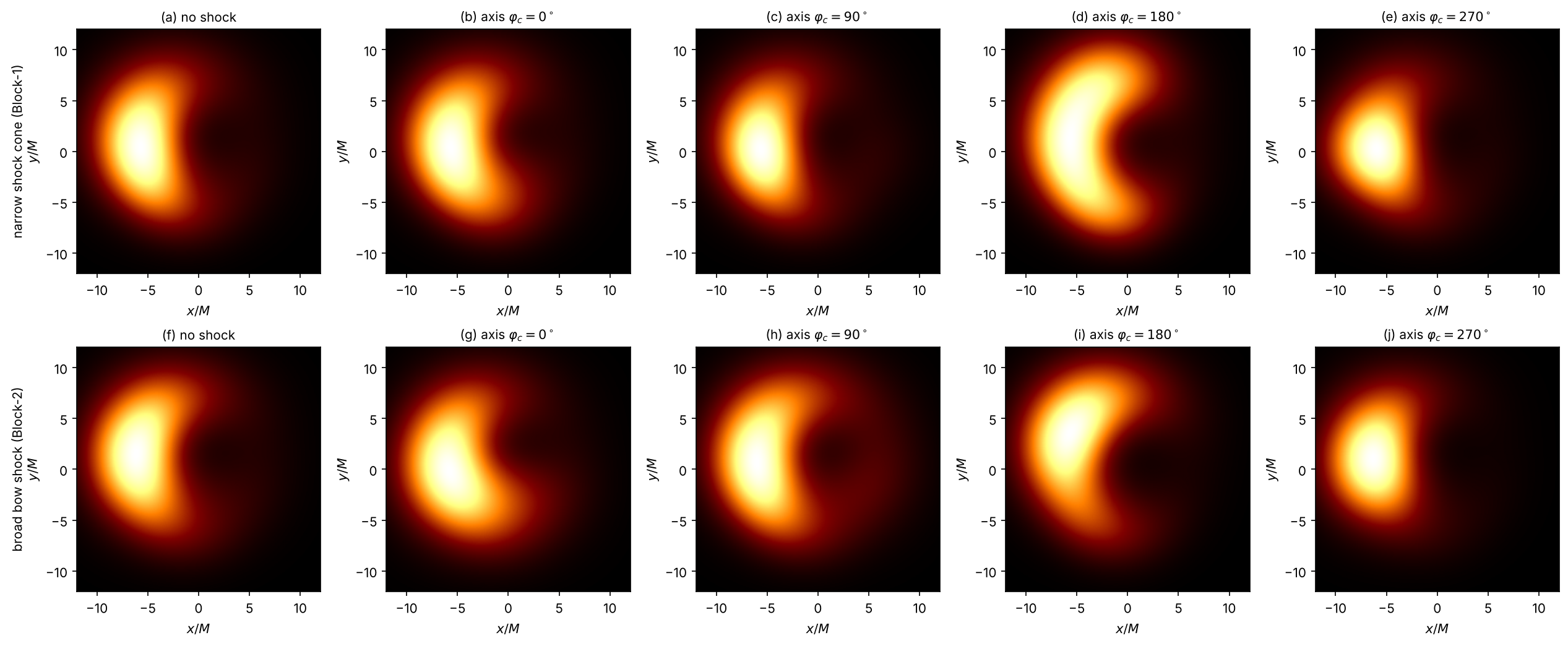}
\caption{{Beam-convolved 230~GHz images ($\theta_o=60^\circ$, toroidal field) including a parametrized shock-density enhancement $n_e\to n_e[1+A\,e^{-\vartheta^2/2w^2}]$, with $A=3$ and $\vartheta$ the angle between the emission point and the shock axis lying in the equatorial plane at azimuth $\varphi_c$ relative to the observer. Top: narrow downstream shock cone ($w=15^\circ$) on the Block-1 metric $(4,1)$. Bottom: broad bow-shock-like structure ($w=45^\circ$) on the Block-2 metric $(1,1.3)$. The first column has no shock; the other columns differ only in the azimuth of the shock axis.}}
\label{fig:cone}
\end{figure*}

{\emph{Connection with shock-cone/bow-shock structures.} To make the link with the GRHD morphology of Ref.~\cite{DonmezMustafaYousafJavedSaidovAtamurotov2025} explicit, we superpose on the fiducial flow a parametrized density enhancement mimicking a shock cone (Fig.~\ref{fig:cone}). Its density contrast, $1+A=4$, corresponds to a strong adiabatic shock, and the opening width $w$ distinguishes the narrow cone of Block-1 ($w=15^\circ$) from the broad bow shock of Block-2 ($w=45^\circ$); the azimuth $\varphi_c$ of the axis is varied relative to the observer. The results answer the three questions raised. (1) \emph{Size and geometry.} The size of the shock structure does not set the scale of the ring: the sharp-image diameter varies by $\lesssim1.5\%$ ($10.59$--$10.78\,M$ for the cone, $11.40$--$11.71\,M$ for the bow shock), because it is fixed by the critical curve. The beam-convolved diameter, however, depends on where the extra emission sits and varies from $9.40$ to $10.90\,M$ (cone) and from $7.67$ to $11.60\,M$ (bow shock). (2) \emph{Brightness.} The shock modulates the flux strongly: relative to the shock-free flow the 230~GHz flux increases by factors $1.08$--$1.48$ for the narrow cone and $1.34$--$2.58$ for the broad bow shock, with the largest enhancement when the dense region lies on the far side of the BH ($\varphi_c=180^\circ$), where it is strongly lensed into the ring. (3) \emph{Viewing azimuth.} Since the structure breaks axisymmetry, all image properties depend on the observer azimuth: the brightness asymmetry along the ring varies between $0.78$ and $0.88$ (cone) and between $0.74$ and $0.91$ (bow shock), and the position angle of the brightest spot shifts by up to $34^\circ$ for the bow shock. The azimuthal dependence is therefore much stronger for the broad Block-2 structure than for the narrow Block-1 cone, which provides an imaging counterpart of the dynamical distinction found in Ref.~\cite{DonmezMustafaYousafJavedSaidovAtamurotov2025}. For a non-stationary shock cone, the same mechanism converts its oscillation into flux and morphology variability.}

\section{Observational Strategy and Discussion}
\label{sec:obs}

The combination of (i) the shadow-radius map $R_{\rm sh}(S_1,S_2)$ of Sec.~\ref{sec:shadow}, (ii) the polarized thick-disk imaging pipeline of Sec.~\ref{sec:disk}, and (iii) the QPO frequency ratios already computed hydrodynamically in Ref.~\cite{DonmezMustafaYousafJavedSaidovAtamurotov2025} constitutes a multi-messenger, horizon-scale test of Lee-Wick gravity: (a) EHT-type shadow measurements of M87* and Sgr A* directly bound $R_{\rm sh}(S_1,S_2)$ (Sec.~\ref{sec:eht-shadow}); (b) polarimetric EHT images \cite{EHT2021VII,EHT2021VIII} constrain the near-horizon magnetic-field topology and, through the disk models of Sec.~\ref{sec:disk}, indirectly constrain $S_1,S_2$ via the induced ring asymmetry and depolarization; and (c) X-ray timing missions such as eXTP \cite{Zhang2025eXTP} and STROBE-X \cite{Froning2024} can test the predicted LFQPO-to-HFQPO transition, providing an independent, accretion-scale handle on the same parameters. Because these three observables respond differently to $S_1$ (exponential damping) and $S_2$ (oscillatory modulation), their combination in principle breaks the degeneracy between the two Lee-Wick couplings that any single observable alone would leave partially unresolved.

{Our results allow a more careful assessment of each channel. (a) With present EHT precision, the shadow size already excludes two strong-coupling bands (Fig.~\ref{fig:heatmap}), but the bound depends on the mass-distance priors and on a Kerr-calibrated ring-to-shadow factor which, as shown in Sec.~\ref{sec:numerics}, changes by $\sim10\%$ in Block-2, and by a similar amount with the disk thickness; a self-consistent analysis would require fitting the Lee-Wick images directly to the interferometric data. Block-1 ($S_1\gtrsim2.5$) remains unconstrained. (b) The polarimetric $\beta_2$ amplitude changes by up to $\sim50\%$ in Block-2, whereas $\angle\beta_2$ at low inclination is set by the field geometry. Since the absolute $|\beta_2|$ of M87* ($0.04$--$0.07$ \cite{EHT2021VIII}) is strongly reduced by Faraday rotation and turbulence, which are absent from our model, a polarimetric test requires a joint fit of plasma and metric parameters; polarization is thus a complementary, not a standalone, probe. (c) Morphology rules out the horizonless corner, since a central brightness depression is observed for M87* and Sgr A*. (d) In the longer term, the photon-ring Lyapunov exponent (Table~\ref{tab:shadow}), accessible to space-VLBI photon-ring interferometry \cite{Johnson2020}, offers a nearly plasma-independent test, with $\gamma$ exceeding $\pi$ by $\gtrsim20\%$, and up to a factor $\simeq3$, in Block-2.}

\section{Conclusions}
\label{sec:conclusions}

We have derived the photon sphere condition and shadow radius of the non-rotating quantum Lee-Wick black hole and shown that the shadow size converges to the Schwarzschild value in the weak-coupling (Block-1) regime but becomes strongly, non-monotonically modified in the strong-coupling (Block-2) regime, in a manner consistent with the shock-cone/bow-shock morphology found in earlier BHL-accretion hydrodynamic simulations. We have mapped the horizon and photon-sphere structure of the full $(S_1,S_2)$ plane, showing that horizonless configurations are regular ultracompact objects with a critical curve but no dark shadow, and that in configurations with two unstable photon spheres the shadow is bounded by the inner critical curve while the outer one produces an additional bright ring. We have further developed the polarized radiative-transfer formalism, using both a phenomenological RIAF-like model and the analytic Hou disk model, needed to image a geometrically thick accretion flow around this spacetime, deliberately bypassing the thin-disk approximation in favor of a description appropriate to horizon-scale, EHT-like targets. {Building on this formalism, we developed a dedicated ray-tracing and polarized-transfer code (planar geodesic integration, thermal synchrotron emission at 230~GHz, exact redshift for a ballistic flow, Walker-Penrose polarization transport) and produced} shadow, blurred, intensity-profile, and polarization images across the Block-1 and Block-2 sequences and as a function of observer inclination. The images confirm quantitatively the $R_{\rm sh}(S_1,S_2)$ trends of Sec.~\ref{sec:shadow}, show that Gaussian-beam convolution at the current EHT resolution blends the primary and higher-order images, and demonstrate that the observer inclination alone generates a strong Doppler crescent even in the absence of spin.

{Our main quantitative results are: (i) Block-1 images are indistinguishable from Schwarzschild to $\lesssim1\%$ in all observables considered; (ii) in Block-2 the ring diameter deviates by up to $\sim14\%$, the photon-ring Lyapunov exponent increases from $\pi$ to $\simeq4$--$9$, and the $\beta_2$ amplitude increases by up to $\sim50\%$ for a mixed field, while $\angle\beta_2$ at low inclination is insensitive to $(S_1,S_2)$; (iii) the $1\sigma$/$2\sigma$ EHT shadow bounds for M87* and Sgr A* exclude two diagonal strong-coupling bands of the $(S_1,S_2)$ plane, while $S_1\gtrsim2.65$ is allowed for all $S_2\in[1,4]$; (iv) shock-cone and bow-shock structures modulate the flux by factors up to $\simeq2.6$ and make the brightness distribution strongly dependent on the viewing azimuth, while leaving the critical-curve size unchanged; and (v) the disk thickness affects the beam-convolved diameter by $\sim10$--$17\%$ and must be fitted jointly with the metric.} A dedicated, full covariant polarized-transfer campaign {including absorption, Faraday effects and GRMHD-based plasma models}, combined with the GRHD-derived shock-cone morphology, is the natural next step toward a complete, quantitative multi-wavelength test of Lee-Wick quantum gravity at the horizon scale.

\bibliography{references}
\bibliographystyle{apsrev4-2}

\end{document}